\documentclass[a4paper,fleqn,usenatbib]{mnras}
\pdfoutput=1

\usepackage{newtxtext,newtxmath}

\usepackage[T1]{fontenc}
\usepackage{ae,aecompl}

\usepackage{graphicx}	
\usepackage{amsmath}	
\usepackage{amssymb}	
\usepackage{bm}

\title[SFR Suppression Due to Local Reionization]{Suppression of Star Formation in Low-Mass Galaxies Caused by the Reionization of their Local Neighborhood}

\author[T. Dawoodbhoy et al.]{
Taha Dawoodbhoy,$^{1}$\thanks{E-mail: \href{mailto:tahad@astro.as.utexas.edu}{tahad@astro.as.utexas.edu}} Paul R. Shapiro,$^{1}$\thanks{E-mail: \href{mailto:shapiro@astro.as.utexas.edu}{shapiro@astro.as.utexas.edu}} Pierre Ocvirk,$^{2}$ Dominique Aubert,$^{2}$\and
Nicolas Gillet,$^{2,3}$ Jun-Hwan Choi,$^{1}$ Ilian T. Iliev,$^{4}$ Romain Teyssier,$^{5}$\and
Gustavo Yepes,$^{6}$ Stefan Gottl{\"o}ber,$^{7}$ Anson D'Aloisio,$^{1,8}$ Hyunbae Park,$^{1,9}$\and
 Yehuda Hoffman,$^{10}$
\\
$^{1}$Department of Astronomy, University of Texas at Austin, Austin, TX 78712-1083, USA\\
$^{2}$Observatoire Astronomique de Strasbourg, Universit\'{e} de Strasbourg, CNRS UMR 7550, 11 rue de l'Universit\'{e}, 67000 Strasbourg, France\\
$^{3}$Scuola Normale Superiore, Piazza dei Cavalieri 7, I-56126 Pisa, Italy\\
$^{4}$Astronomy Center, Department of Physics \& Astronomy, Pevensey II Building, University of Sussex, Falmer, Brighton BN1 9QH, UK \\
$^{5}$Institute for Theoretical Physics, University of Z\"{u}rich, Winterthurerstrasse 190, CH-8057  Z\"{u}rich, Switzerland\\
$^6$Departamento de F\'{\i}sica Te\'{o}rica and CIAFF, M\'{o}dulo 15 Universidad Aut\'{o}noma de Madrid, 28049, Madrid, Spain\\
$^{7}$Leibniz-Institute f\"{u}r Astrophysik Potsdam (AIP), An der Sternwarte 16, D-14482 Potsdam, Germany\\
$^{8}$Department of Physics \& Astronomy, University of California, Riverside, CA 92521, USA\\
$^{9}$Korea Astronomy and Space Science Institute, Daejeon, 305-348, Korea\\
$^{10}$Racah Institute of Physics, Hebrew University, Jerusalem 91904, Israel}

\date{Accepted XXX. Received YYY; in original form ZZZ}

\pubyear{2018}

\begin{document}
\label{firstpage}
\pagerange{\pageref{firstpage}--\pageref{lastpage}}
\maketitle

\begin{abstract}
Photoheating associated with reionization suppressed star formation in low-mass galaxies. Reionization was inhomogeneous, however, affecting different regions at different times. To establish the causal connection between reionization and suppression, we must take this local variation into account. We analyze the results of CoDa (`Cosmic Dawn') I, the first fully-coupled radiation-hydrodynamical simulation of reionization and galaxy formation in the Local Universe, in a volume large enough to model reionization globally but with enough resolving power to follow all atomic-cooling galactic halos in that volume. For every halo identified at a given time, we find the redshift at which the surrounding IGM reionized, along with its instantaneous star formation rate (`SFR') and baryonic gas-to-dark matter ratio ($M_\text{gas}/M_\textsc{dm}$). The average SFR {\it per halo} with $M < 10^9 \text{ M}_\odot$ was steady in regions not yet reionized, but declined sharply following local reionization. For $M > 10^{10} \text{ M}_\odot$, this SFR continued through local reionization, increasing with time, instead. For $10^9 < M < 10^{10} \text{ M}_\odot$, the SFR generally increased modestly through reionization, followed by a modest decline. In general, halo SFRs were higher for regions that reionized earlier. A similar pattern was found for $M_\text{gas}/M_\textsc{dm}$, which declined sharply following local reionization for $M < 10^9 \text{ M}_\odot$. Local reionization time correlates with local matter overdensity, which determines the local rates of structure formation and ionizing photon consumption. The earliest patches to develop structure and reionize ultimately produced more stars than they needed to finish and maintain their own reionization, exporting their `surplus' starlight to help reionize regions that developed structure later.
\end{abstract}




\begin{figure*}
    \centering
    \includegraphics[width=\textwidth]{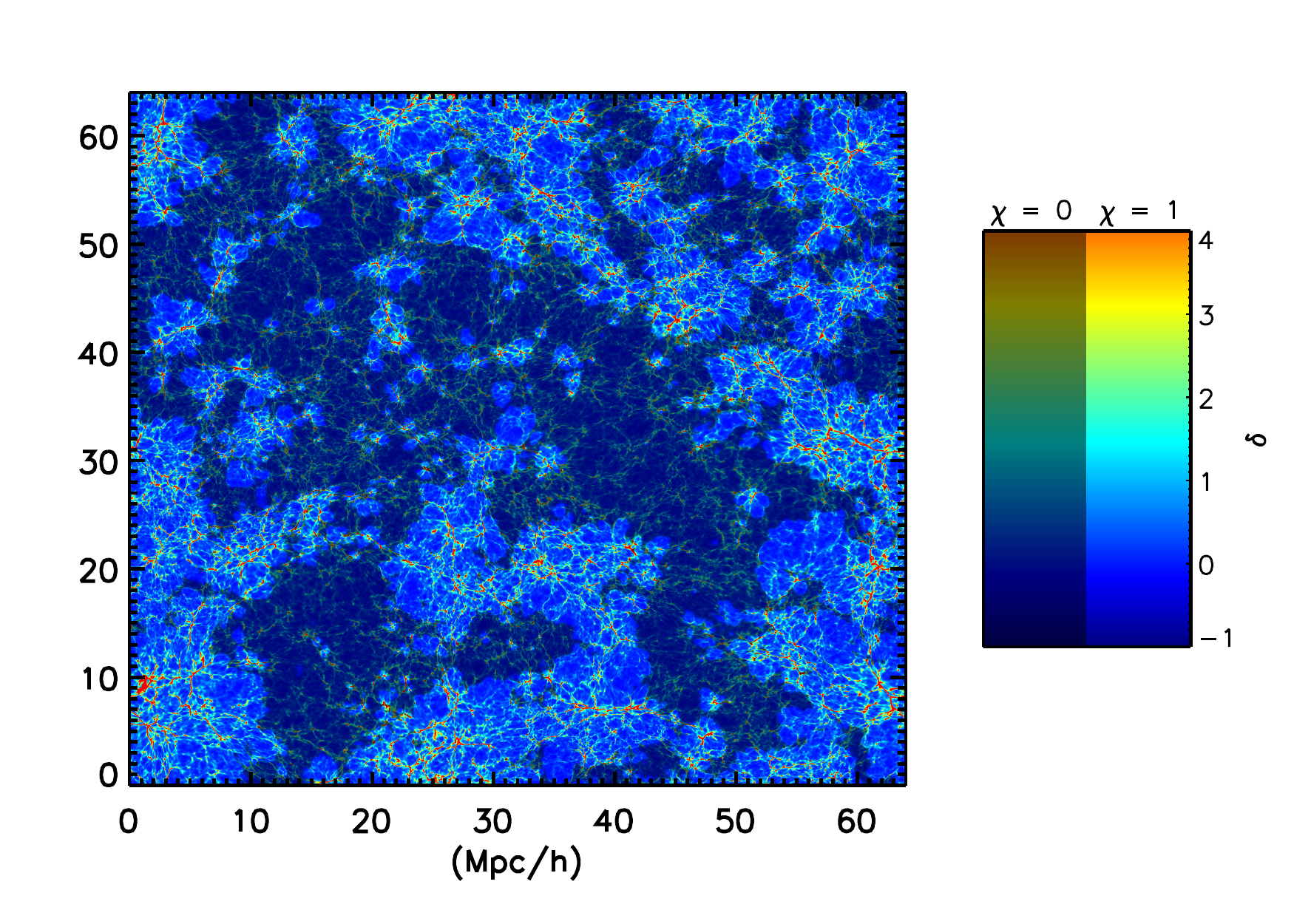}
    \caption{A visualization of a slice through the CoDa simulation when half of the IGM was ionized ($z=5.44$) showing the cosmic web and the `patchiness' of reionization. Gas overdensity, $\delta$, increases from blue to red. The color brightness depends on the ionized fraction, $\chi$, with the darkest regions corresponding to completely neutral gas and the brightest regions corresponding to completely ionized gas. These extremes are indicated in the legend to the right.}
    \label{fig:vis}
\end{figure*}

\section{Introduction}

When the first galaxies born from the growth of density fluctuations in the early universe formed stars, some of their UV starlight escaped into the surrounding intergalactic medium (IGM) and photoionized H atoms there, heating the gas to temperatures $T\sim10^4$ K, to start the Epoch of Reionization (EOR). Over time, as more galaxies formed more stars, the H II regions surrounding them grew and overlapped, led by the expansion of highly supersonic ionization fronts that raced ahead of the hydrodynamical back-reaction of the IGM to its photoheating. Inside these H II regions, however, that back-reaction was inescapable. Current evidence suggests that the entire IGM was transformed into an H II region in this way by a redshift of $z\sim6$, less than a billion years after the Big Bang. By the end of the EOR, therefore, the hydrodynamical consequences of reionization were everywhere. Reionization was `patchy', however -- regions with the most advanced structure formation were reionized first, while those with the slowest structure formation were ionized last -- so its effects were inhomogeneous and asynchronous, too (see Fig.~\ref{fig:vis}).

The EOR is believed to have had a substantial impact on galaxy formation. For example, the photoheating of gas in and around dwarf galaxies is often proposed as a solution to the `missing satellites problem' -- the over-abundance of dwarf satellites in the Local Group (`LG') predicted by N-body simulations of the $\Lambda$CDM model as compared with observations -- by the suppression of their star formation that would result from it \citep[][]{Klypin,Moore,Bullock,Somerville,BensonA,BensonB,Benson3,BBK17}. 
As described by \citet[]{Shapiro}, the heating of the IGM introduced pressure forces which opposed the growth of linear perturbations in the baryonic component by gravitational instability -- sometimes referred to as `Jeans-mass filtering' -- thereby reducing the rate of collapse of baryons out of the IGM into dark matter halos.
In that case, dark-matter-dominated halos would still have formed, but without their fair share of baryons. For halos that formed before the surrounding IGM was reionized, pressure forces in the IGM {\it after} its reionization would also have prevented further accretion of baryons onto the pre-existing galaxy if its mass was below a value comparable to the intergalactic Jeans mass.
So, while such galaxies may also lose gas from various internal processes, such as supernovae, radiation pressure, stellar winds, interactions with larger galaxies, etc. (and photoheating of their interstellar medium by external radiation during reionization may also contribute to gas loss), they would not have been able to accrete much of it back. Both of these effects worked to suppress star formation preferentially in low-mass galaxies, and thus potentially rendered Local Group dwarfs invisible. The same effect would have suppressed star formation in such low-mass galaxies everywhere else, as well, thereby affecting the history of reionization itself, both {\it locally} and {\it globally}.

The characteristic value of the mass scale below which halos are suppressed by this reionization feedback in the CDM universe has been estimated by various approximations over the years, including analytical and semi-analytical estimates \citep[e.g.][]{Rees86,CR86,Efs}, linear perturbation theory combined with a thermal and ionization history of the reionizing IGM \citep[e.g.][]{Shapiro}, equilibrium models \citep[e.g.][]{Ik86,KBS97}, 1-D numerical hydrodynamics with and without radiative transfer \citep[e.g.][]{UIk84,TW96,HTL96,Kitayama_et_al_2000,Dij04,SM13}, and optically-thin, 3-D N-body and hydrodynamics simulation  \citep[e.g.][]{CO92,QKE96,NS97,Gnedin00,Hoeft_et_al_06,OGT08}. 
Most recent 3-D hydro+N-body simulations of galaxy formation still approximate the effects of reionization feedback by assuming a uniform UV background switches on at a predetermined `reionization redshift', without radiative transfer or else with only an approximation that shields gas above some density \citep[e.g.][]{Kuhlen,Nelson_et_al_2015,Fitts_et_al_2018}.
A range of halo mass scales (or virial velocities) subject to suppression have been reported, with a typical value below which either star formation or the accretion of intergalactic gas are suppressed of order $10^9\text{ M}_\odot$.  

The back-reaction of this suppression on the history of reionization and its large-scale inhomogeneity was considered by \citet[]{Iliev07}. They post-processed an N-body simulation with radiative transfer and non-equilibrium ionization and assumed that star formation in galactic halos below $10^9 \text{ M}_\odot$ was suppressed if they were located within an ionized patch of the IGM. \citet[]{Iliev07} found that reionization `self-regulated' because of this feedback. The suppressible low-mass atomic-cooling halos (LMACHs) started reionization, limited by the growing fraction of the volume inside H II regions, but high-mass atomic-cooling halos (HMACHs) dominated its completion.
To do this right, however, one must ultimately simulate galaxy formation and reionization {\it together}, with fully-coupled hydrodynamics and radiative transfer (RT), and with enough dynamic range to resolve this mass scale, a computational challenge that has only recently been met. In what follows, we will describe one such simulation and apply it to demonstrate the causal connection between reionization and the suppression of individual galaxies, and its dependence on halo mass.
We will compare this work with other recent simulations, as well.

We focus here on the mass range of atomic-cooling halos, believed to be the dominant contributors to reionization. Minihalos (MHs) -- halos with $M<10^8\text{ M}_\odot$ and virial temperatures $T<10^4$ K -- also form and are believed to be the sites of the first star formation in the $\Lambda$CDM universe. However, their contribution to reionization is sub-dominant. MH star formation is only possible if H$_2$ molecules form and cool their gas well below the virial temperature, but dissociating radiation in the Lyman-Werner (LW) bands below 13.6 eV can prevent this if the intensity is high enough. The LW background from the same stars that contribute to reionization rises above this threshold long before the end of the EOR \citep{HRL97,Ahn12}. In addition, MHs are also vulnerable to photo-evaporation by intergalactic I-fronts during the EOR, so their gas content is expelled if they are located inside the H II regions of the IGM \citep{Shapiro04,Ilian05,Park16}.

In \S~\ref{Simulation}, we discuss the requirements for simulating global, inhomogeneous reionization self-consistently and with good enough statistics to sample its inhomogeneous feedback fairly. We then introduce the CoDa\footnote{
Henceforth, we will refer to the CoDa simulation as `CoDa~I' to distinguish it from other simulations to come in the series, including CoDa I-AMR recently described in \citet[]{codaIamr}.}
simulation, which meets these requirements, and summarize its globally-averaged reionization and suppression results. In \S~\ref{local}, CoDa I results are analyzed to show that the {\it inhomogeneous} nature of reionization leads to {\it inhomogeneous} feedback effects on galactic star formation rates and baryon gas content. Our results are summarized and discussed in \S~\ref{conclusion}.

\section{CoDa (Cosmic Dawn) I: Simulating the EOR with Fully-Coupled Radiation-Hydrodynamics}
\label{Simulation}
\subsection{What are the minimum simulation requirements for this study?}
In order to determine the impact of reionization on halo properties like their star formation rate (SFR) and gas-to-dark matter ratio ($M_\text{gas}/M_\textsc{dm}$), it is necessary to consider the history of reionization {\it local} to each halo, which is different for halos in different locations. To simulate this, we must be able to represent a subvolume of the universe large enough to sample the full range of experiences of individual halos while following the spatial inhomogeneity of reionization on large-scales, with enough resolution to track the formation of all star-forming halos in that volume, as well as the mutual feedback of these halos and reionization on each other, with fully-coupled radiation-hydrodynamics from cosmological initial conditions. To sample reionization inhomogeneity fairly, a comoving volume of order 100 Mpc on a side is required \citep{iliev2014}. Resolving all atomic-cooling halos (i.e. those with masses $M \gtrsim 10^8 \text{ M}_\odot$) with hundreds of N-body particles requires a particle mass smaller than $10^6 \text{ M}_\odot$, implying as many as $\sim 10^{11}$ particles for such a large volume.

Our goal is to quantify the dependence of the SFRs and halo gas-to-dark matter ratios of individual galaxies on their local reionization histories. For that, we must be able to subdivide the total number of halos identified at each redshift in this simulated volume into bins of cosmic time (or redshift) during which their surroundings were reionized, and then subdivide these bins further into bins of common halo mass, with enough halos in each bin to enable us to average over the halo-to-halo scatter in their SFRs and gas fractions. This requirement is yet another reason the simulation must have a very large volume. 

\subsection{CoDa I fits the bill}
The combination of such a large volume and high resolution in a fully-coupled radiation-hydro simulation of reionization has only recently become computationally feasible.
{\it{CoDa}}
(Cosmic Dawn) I, described in \citet[]{Ocvirk}, is the first such simulation to meet all of these requirements.
There is a fundamental mismatch between the small time steps required to resolve radiative transfer and non-equilibrium ionization in the presence of highly supersonic ionization fronts and the larger time steps which are sufficient to resolve the mass motions associated with pressure and gravity forces. Forcing the simulation to use the smaller of these two step sizes for both is computationally prohibitive and wasteful. CoDa I is based on the hybrid CPU-GPU code, \textsc{ramses-cudaton}, which overcame this problem by performing hundreds of RT steps on the GPUs per gravito-hydrodynamics step on the CPUs, on each of 8192 nodes of the CPU-GPU hybrid, massively-parallel supercomputer Titan at OLCF.
It has a resolution of $4096^3$ cells for hydrodynamics and radiative transfer and the same number of dark matter particles (each with a mass of $3.49 \times 10^5$ M$_\odot$) in a comoving box 91 cMpc on a side, and is simulated down to a redshift of 4.23. 

The initial conditions (I.C.'s) for CoDa I were a realization of Gaussian random noise density fluctuations in the $\Lambda$CDM universe, assuming cosmological parameters from the case ``WMAP + BAO + SN'' in \citet[]{Hinshaw09} (i.e. $\Omega_m=0.279$, $\Omega_\text{bary}=0.046$, $\Omega_\Lambda=0.721$, $h=0.7$, $\sigma_8=0.817$, $n_s=0.96$). They were a `constrained realization' produced by the CLUES (``Constrained Local UniversE Simulations'') consortium \citep[][]{clues}, derived from observations of galaxies in our Local Universe, and designed to reproduce its observed features when evolved cosmologically over time to the present (e.g. 
the Local Group, containing structures resembling the Milky Way and M31, and the Virgo and Fornax clusters). This allows us to compare features of CoDa I with Local Group observations directly, in a volume large enough to model global reionization, too \citep[see][]{Ocvirk}.~\footnote{Towards this end, these same CLUES I.C.'s were recently used by another CoDa simulation, CoDa I-AMR, based on the hybrid CPU-GPU, AMR radiation-hydrodynamics code EMMA, as described by \citet[]{codaIamr}, where objects in the Local Universe at $z=0$ were matched to their progenitors during the EOR at $z\geq6$, to study the inhomogeneous reionization times of present-day galaxies.}


\subsection{Results from CoDa I on Global Reionization and the Suppression of Star Formation}
\citet[]{Ocvirk} present a wide range of results from CoDa I, but here we will focus only on those pertaining to the suppression of star formation in low-mass galaxies. We note that the EOR ended somewhat later in the CoDa I simulation than estimated from current observations, with globally-averaged, volume-weighted ionized fraction $X_\text{ion}\simeq0.9$ at $z\simeq4.9$ and $1-X_\text{ion}\simeq10^{-4}$ at $z\simeq4.6$. The efficiency parameters chosen for our subgrid star-formation algorithm, calibrated by a series of small-box simulations, were, as it turned out, a bit lower than required to finish reionization by $z=6$. Fortunately, a simple temporal re-scaling, $z \rightarrow 1.3z$, serves to align the data from CoDa I with many independent observational constraints, as shown in \citet[]{Ocvirk}, so it is useful to use this re-scaling when relating CoDa I to those observables. However, the results we present here are not required to be related to observations made at any specific redshift. Rather, we focus on the relative difference in redshift of reionization (see Section~\ref{local}) between different patches, so we will simply use the original redshift values of the CoDa I simulation, not the re-scaled ones.

\begin{figure}
    \centering
    \includegraphics[width=\columnwidth]{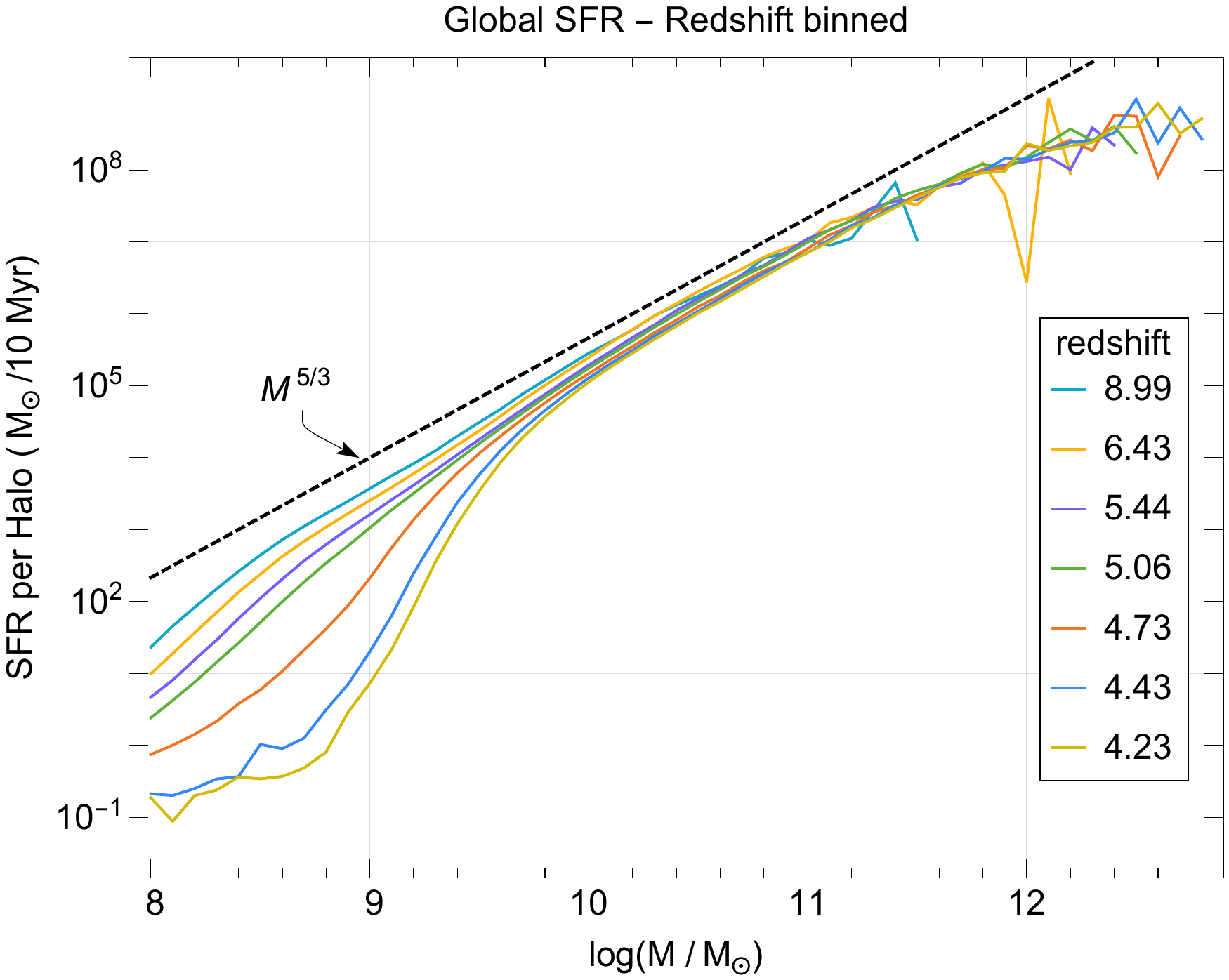}
    \caption{Instantaneous SFR per halo versus instantaneous halo mass in CoDa I, averaged over the entire simulation volume, for various redshifts. Notice the sharp suppression at low mass at later redshifts.}
    \label{fig:coda7}
\end{figure}
\begin{figure}
    \centering
    \includegraphics[width=\columnwidth]{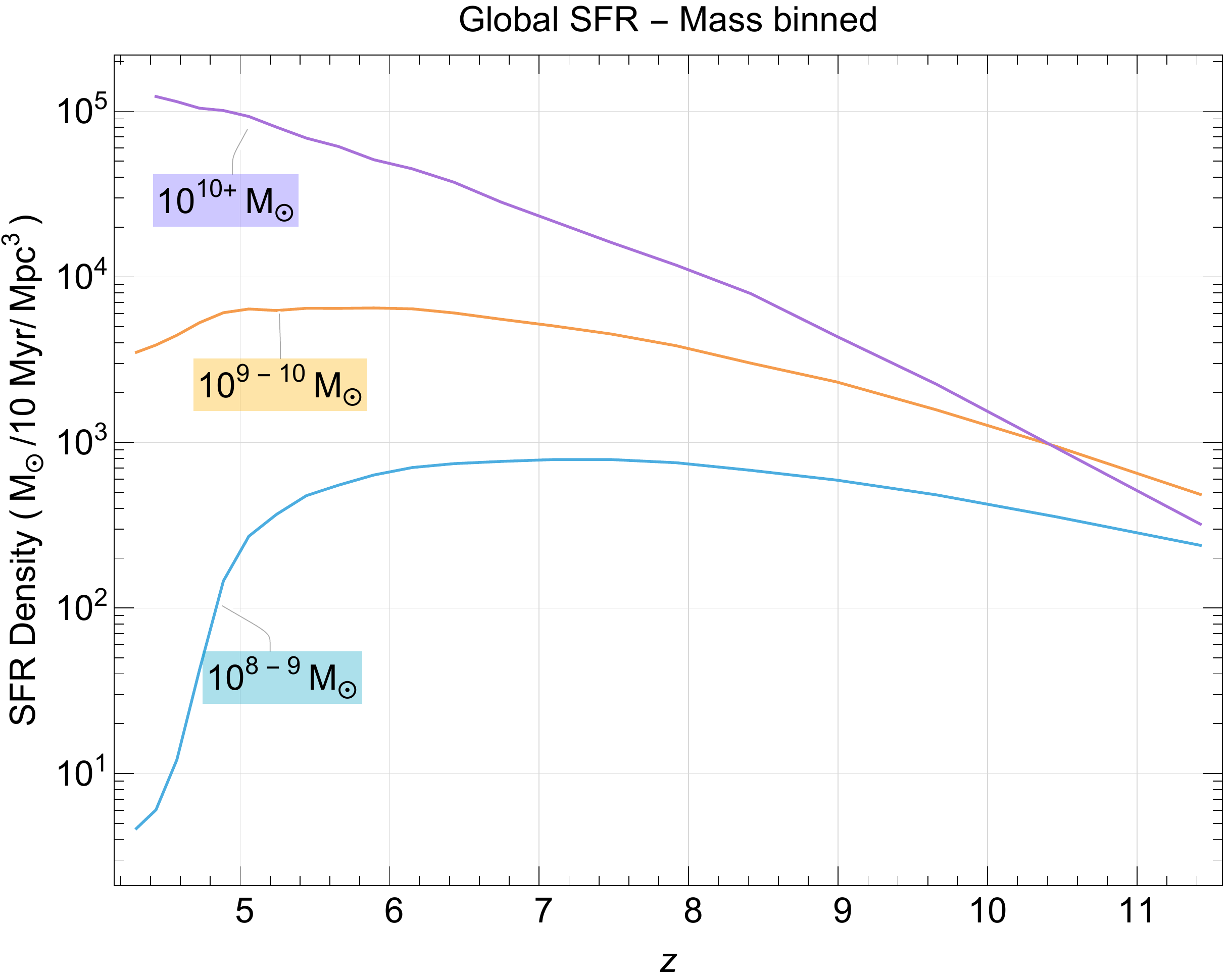}
    \caption{Instantaneous SFR density versus redshift, globally-averaged, for low, intermediate, and high halo-mass bins in the CoDa I simulation.}
    \label{fig:coda9}
\end{figure}

The results for CoDa I, when averaged over the entire volume, showed that,
at early times, the SFR per halo for halos of all masses\footnote{As described in \citet[]{Ocvirk}, halos are identified in the N-body data by a Friends-of-Friends algorithm. Therefore, the halo mass to which we refer here is the mass in dark matter.} can be approximated by a power law:
\begin{equation}
    \text{SFR} \sim M^{5/3}
    \label{eq:M53}
\end{equation}
At later times, massive halos ($M \gtrsim 3\times 10^{9}$ M$_\odot$) continue to follow this relationship, while lower mass halos appear suppressed relative to it. The strength of this suppression is monotonically related to the mass of the halo -- lower mass halos feel stronger suppression. By $z=4.23$, SFRs in halos with masses between $10^{8}$ and $10^9$ M$_\odot$ were suppressed by up to a factor of $10^2$ relative to $z=6.43$ (Fig.~\ref{fig:coda7}). The contribution to the global SFR density by these low-mass halos decreased sharply at late times. A weaker suppression was seen for halos between $10^{9}$ and $10^{10}$ M$_\odot$, and no suppression was seen for halos with $M \gtrsim 10^{10}$ M$_\odot$ (Fig.~\ref{fig:coda9}).

A convergence test of the suppression mass-scale (i.e. the mass below which halos are suppressed) is described in \citet[]{Ocvirk}. A series of small-box versions of the CoDa I simulation were run to determine how the suppression mass-scale depends on resolution. Each of these simulations used $512^3$ cells, but had varying box sizes ranging from 4 to 32 $h^{-1}$ Mpc on a side. The 8 $h^{-1}$ Mpc version had the same resolution as CoDa I. The suppression mass-scale was found to depend on resolution in general, but the 4 $h^{-1}$ Mpc test run yielded the same suppression mass-scale as the 8 $h^{-1}$ Mpc test run, indicating convergence. In addition, a second version of the 8 $h^{-1}$ Mpc simulation was run, but without radiative transfer, so that the only sources of feedback were supernovae. This version displayed far weaker suppression in low-mass halos relative to the version that included radiative transfer, which suggests that reionization feedback is the primary driver of suppression in the CoDa I simulation.

While CoDa I is a first of its kind in being a fully-coupled radiation-hydro simulation with enough volume and enough resolution to model large-scale reionization and galaxy formation simultaneously, both {\it locally} and {\it globally}, other recent simulations with some, but not all, of these characteristics have also been reported with implications for the issue of reionization suppression, as well.
In comparison with this literature, our results are consistent with those of \citet[]{Kuhlen}, \citet[]{HS13}, and \citet[]{Pawlik}.
Low-mass halo suppression was also seen by the Renaissance simulations of \citet{Renaissance}, starting from a simulated volume 10 times smaller than CoDa I, then re-simulating three special `zoom-in' subregions -- `zoom-ins' with significantly better mass and spatial resolution than CoDa I's but in volumes that were thousands of times smaller. 
However, they report, not the suppression of the SFR per halo vs halo mass, but rather the fraction of halos that experienced {\it any} star formation in the last 20 Myr of their simulation vs halo mass.
They found the mass-scale below which this fraction is reduced (in their zoom-in regions) to be about an order of magnitude lower than our SFR suppression mass-scale (averaged over our full volume).
These results do not contradict each other, however, since our $\sim 10^9\text{ M}_\odot$ halos have non-zero SFRs even after they are suppressed.
Furthermore, their zoom-in simulations were not allowed to react back on the coarser-grained background simulation or influence it in any way, however, so it is difficult to compare these results directly with ours.
Low-mass halo suppression was not apparent in the simulations by \citet[]{Jaacks}, but their low-mass halo results consisted of halos with {\it stellar} masses in the range $10^{6.8}\text{ M}_\odot < M_\star < 10^8\text{ M}_\odot$. Assuming typical stellar mass fractions of $10^{-2}$ -- $10^{-3}$, this stellar mass range could very well consist mostly of halos that are massive enough to avoid suppression according to our results. Finally, \citet[]{croc2} found that reionization feedback had a very modest impact on low-mass galaxies in the CROC simulation, but a direct comparison with CoDa I is difficult since they studied the UV luminosity function but do not report the SFR-halo mass relation. The difference between their results and ours is discussed further in \citet[]{Ocvirk}.

The globally-averaged results presented in this section demonstrate a correlation between the end of global reionization and SFR suppression in low-mass halos. Thanks to its large volume and dynamic range, CoDa I makes it possible to go well beyond these globally-averaged results, however. Since reionization was inhomogeneous, so must the effects of reionization suppression have been. By studying the inhomogeneity of reionization and its impact on individual halo properties, we will make a more compelling case for the causal relationship between reionization feedback and suppression and its environmental dependence. We do this by monitoring the direct and immediate effects of reionization feedback in many small patches, each reionized at a different time and place (with different overdensities, halo mass functions, halo gas fractions, etc.). In the following sections, we will show that the suppression of low-mass halos universally follows {\it local} reionization, regardless of when or where it occurs.

\section{Local Reionization}
\label{local}
\begin{figure}
    \centering
    \includegraphics[width=\columnwidth]{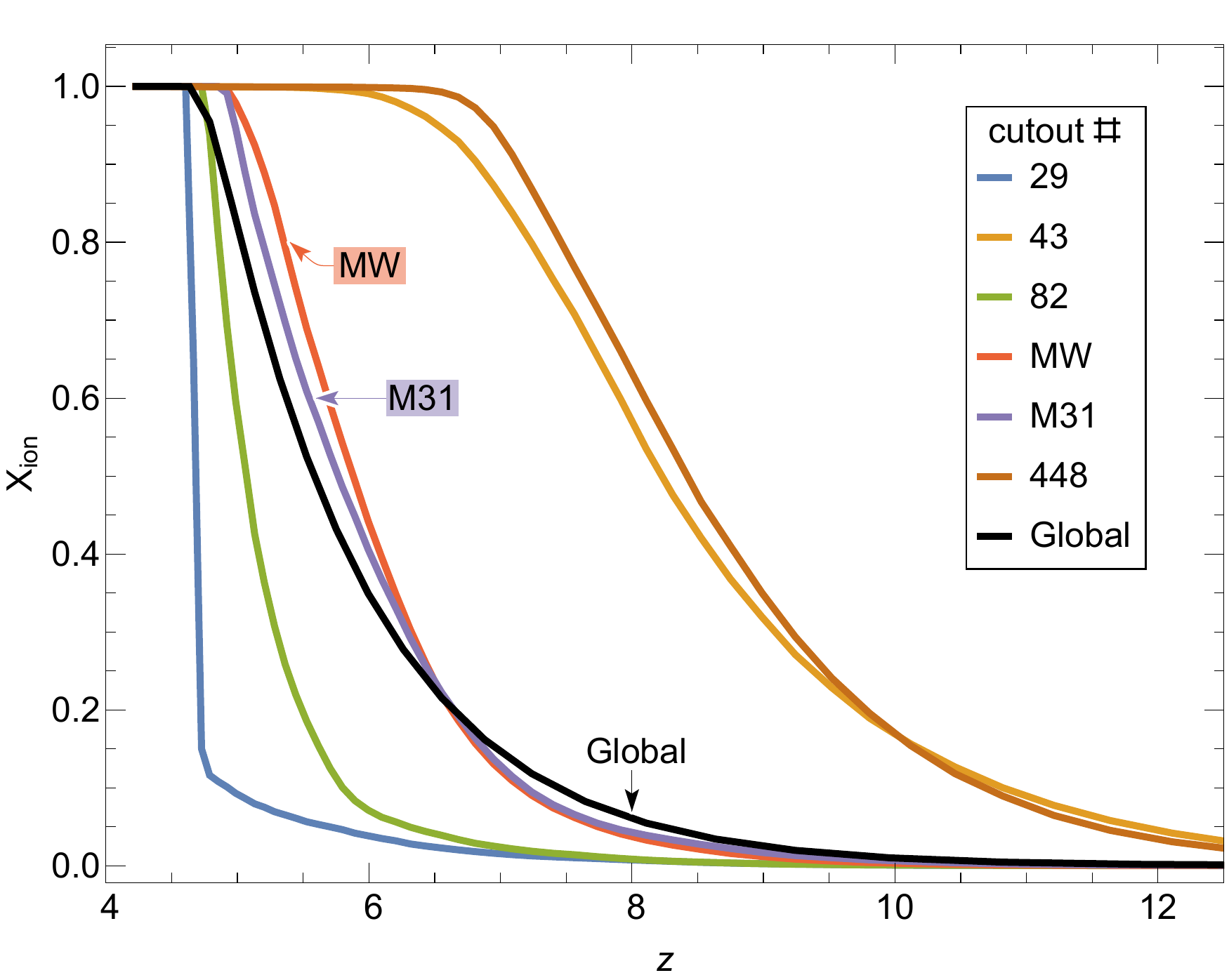}
    \caption{Ionization histories (volume-weighted ionized fraction $X_\text{ion}$ vs redshift) for a selection of different cutouts of the CoDa I simulation, along with the globally-averaged history, for comparison. These cutouts are each 4 $h^{-1}$ Mpc on a side and centered on some objects of interest, such as the Milky Way (`MW') and M31, as identified by evolving the same I.C.'s forward in time to $z=0$ by a separate high-resolution N-body simulation, \mbox{CoDa I-DM2048}.}
    \label{fig:xion}
\end{figure}

The history of reionization was different in different places, as can be seen in Fig.~\ref{fig:xion}. This plot shows the ionization histories of a selection of different CoDa I simulation `cutouts' -- regions of the simulation that are 4 $h^{-1}$ Mpc on a side and centered on some object of interest, such as the Milky Way (`MW'), as identified by evolving the same CLUES constrained initial conditions to $z=0$ in a separate N-body simulation -- the CoDa I-DM2048 \textsc{gadget} simulation, as described in \citet[]{Ocvirk} -- and tracing that region's Lagrangian mass at $z=0$ back to its location during the EOR. As can be seen, these regions not only reionized at different times, but the nature of their reionization was different, as well. Cutouts 43 and 448 (the two right-most curves) represent overdense regions, with mean overdensities $\delta \equiv \rho/\bar{\rho}-1=$ 0.64 and 0.74, respectively, at $z=11.4$. They started reionizing at an early redshift, and did so gradually, reaching volume-weighted ionized fraction $X_\text{ion}=0.9$ at $z_{re} = 6.8$ and 7.1, respectively. Cutouts 29 and 82 (the two left-most curves), on the other hand, represent underdense regions, with mean overdensities of -0.25 and -0.12, respectively, at $z=11.4$. They did not start reionizing until rather late, and did so abruptly, reaching an ionized fraction of 1 at an almost vertical slope (at $z_{re}=4.6$ and 4.8, respectively). 
Further analysis suggests that these different local reionization histories reflect the relative importance of the contributions to local reionization from sources which are either `internal' or `external' to each local subvolume. Histories like those of cutouts 43 and 448, which begin early, are gradual, and approach the end of reionization ($X_\text{ion}\rightarrow1$) with a slope of zero, correspond to regions which were `internally' reionized. At the other extreme, those like cutout 29, which start late, rise abruptly from $X_\text{ion}\sim0.1$ to $X_\text{ion}\sim1$ almost instantly, approaching the end with an infinite slope, correspond to regions which are `externally' reionized. In that case, the rise-time is so short because the externally-driven I-front swept across the comoving distance of $4h^{-1}$ Mpc in too little time for the plot to resolve it visually on the $z$-axis. The histories for the MW and M31 cutout volumes in the Local Group are intermediate between these two extremes, consistent with a significant, perhaps even dominant, contribution from `internal' sources but a non-negligible contribution from `external' sources, too. They finished reionization a bit earlier (at $z_{re}=5.2$ and 5.1, respectively, for $X_\text{ion}=0.9$) than the globally-averaged history (for which $X_\text{ion}=0.9$ at $z_{re}=4.9$) but are qualitatively similar to the latter. The mean densities of the MW and M31 cutouts are closer to the cosmic mean density as well, with overdensities of 0.16 and 0.09, respectively, at $z=11.4$. We discuss the correspondence between local overdensity and reionization timing further in \S~\ref{HMFandOD}.

\begin{figure}
    \centering
    \includegraphics[width=\columnwidth]{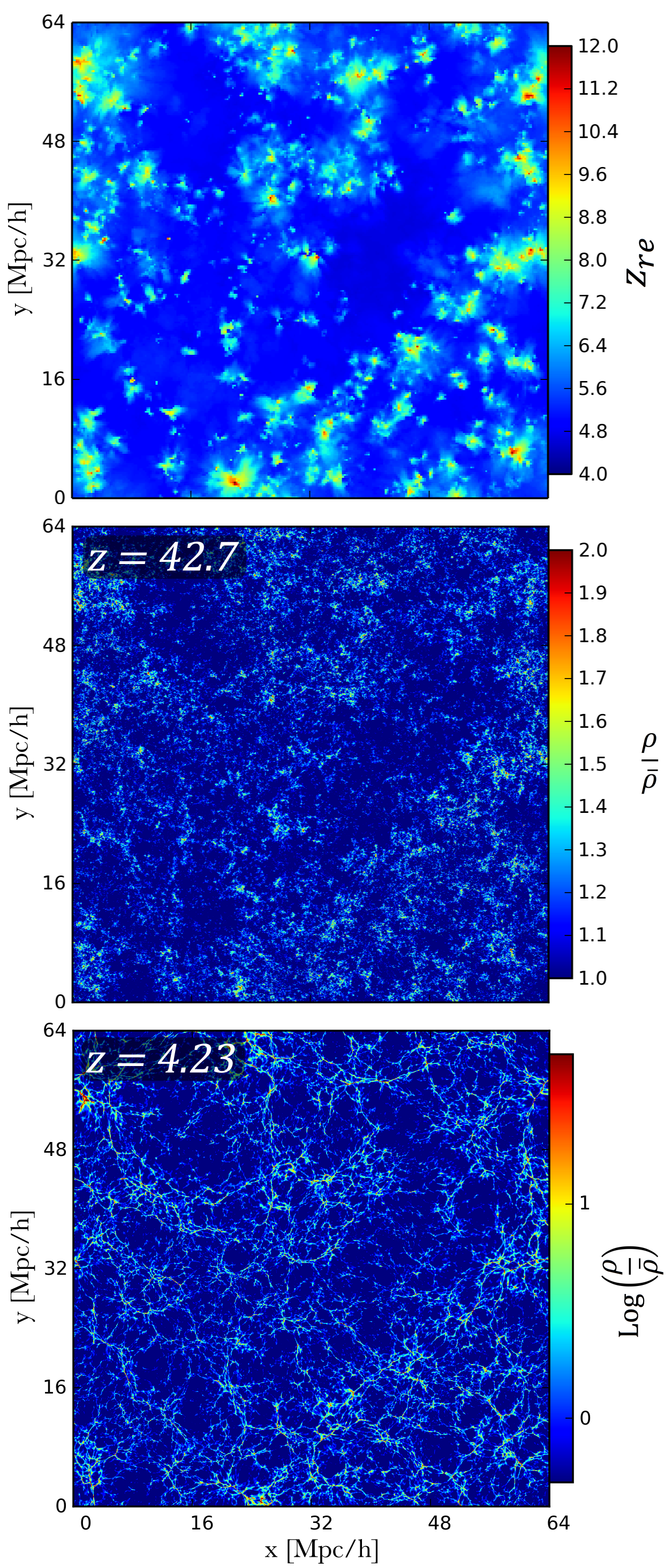}
    \caption{A two-dimensional slice of the CoDa I volume showing the reionization redshift field (top) and the density fields at $z=42.7$ (middle) and $z=4.23$ (bottom), respectively. The $z_{re}$ field shows the rarity of early-reionized regions and the abundance of late-reionized regions (87\% of cells reionized after $z=6.61$). The density fields illustrate that these regions correspond to overdensities and voids, respectively. Each slice is one cell thick for its $256^3$ (top) and $2048^3$ (middle, bottom) grids, respectively.}
    \label{fig:zre}
\end{figure}

As illustrated by these different outcomes, which correlate with the large-scale
patchiness of the EOR, we are required to treat every point in space as having a unique reionization history. We define a `reionization redshift field'\footnote{This reionization redshift field was also discussed by \citet[]{Battaglia}}, $z_{re}(\vec{x})$, to be the redshift at which a small volume surrounding position $\vec{x}$ first reached ionized fraction $X_\text{ion}= 0.9$. Because of the intimate connection between structure formation and reionization, this field resembles the underlying cosmic web of dark matter, galaxies, and the IGM (see Fig.~\ref{fig:zre}): the highest density peaks in the early universe reionized first, while voids were the last to reionize, as we will show in the following sections.

\subsection{Methodology}
\subsubsection{Generating the Reionization Redshift Field}
To track the effect of reionization on low-mass halos, we use the instantaneous locations of halos identified in the CoDa I simulation N-body data, as described in \citet[]{Ocvirk}, and assign to each halo the reionization redshift of its surrounding IGM.
A natural way to do this is to find $z_{re}$ for the computational cell in which the halo resides. However, CoDa I is too finely resolved for this approach to work as is, since halos can be at least several times larger than individual cells. Hence, for the purpose of calculating the reionization redshift field, we smoothed the ionized fraction data from the underlying grid of $4096^3$ cells to a coarsened resolution of $256^3$ cells, each 0.357 cMpc on a side. At this resolution, halos typically make up around 1\% of the cell's volume, so information about these cells mostly reflects characteristics of the IGM. Therefore, the reionization redshift of the IGM surrounding a halo is simply the reionization redshift of the coarse-grained cell in which that halo is located. Additionally, for cells this large, we can neglect the possibility that halos may migrate across cell boundaries on time-scales that would affect our results.

\subsubsection{Extracting the Instantaneous Star Formation Rate per Halo}
To obtain the instantaneous SFR for each halo we must identify all the star particles which are currently associated with that halo. The simulation contains star particles that were produced in one of the original $4096^3$ computational cells whenever the fractional overdensity of the gas in that cell exceeded 50 and the gas temperature was less than $2\times 10^4$ K. 
If gas is radiatively cooling from a higher temperature than this, on its way to becoming gravitationally unstable and forming stars, it will generally proceed to cool below $2\times10^4$ K and be counted later as star-forming gas when it does so. If it is photoionized, however, it will generally not continue to cool below this temperature unless it can self-shield and recombine. Until it does so, however, it is not counted as star-forming gas.
Star particles are produced at a rate of 
\begin{equation}
   \dot{\rho}_\star = \epsilon_\star \frac{\rho_{\text{gas}}}{t_\text{ff}}, 
\end{equation}
where $\epsilon_\star$ is the star formation efficiency ($\epsilon_\star=0.01$), $\rho_{\text{gas}}$ is the gas density, and $t_\text{ff}$ is the free fall time of the gaseous component given by $t_\text{ff}=\sqrt{3\pi/32G\rho_{\text{gas}}}$. The mass of the star particle depends on the gas density of the cell, but is always a multiple of a fixed elementary value, 3549 M$_\odot$. This value was chosen to be small enough to allow effective sampling of low-mass galaxy SFRs, while being large enough to allow us to neglect stochastic variations in the number of high-mass stars represented by each star particle. We assumed that 10\% of the mass of each star particle is in the form of high-mass stars that end their lives as supernovae within 10 Myr. So 10 Myr after the star particle is created, its mass is adjusted to 90\% of its original value. Star particles are assigned to a halo if they are located within a sphere centered on the halo's center of mass, of radius $R_{200}$, the radius within which the volume-averaged dark matter density is 200 times the mean. The instantaneous SFR for each halo, in units of M$_\odot(10$ Myr$)^{-1}$, is simply the total mass of all star particles assigned to the halo that are less than 10 Myr old. With these two pieces of information, we can monitor how the SFRs of halos of a given mass change over time in relation to their {\it local} reionization. 

It is worth mentioning that we refrain from analyzing halos with masses less than $10^{8} \text{  M}_{\odot}$. 
While the Friends-of-Friends (FOF) halo finder used to identify halos in the \mbox{N-body} data of CoDa I did identify halos less massive than this, their number density and identification are not secure.
There is a discrepancy at the smaller masses in the halo mass function between CoDa I and CoDa I-DM2048, a \textsc{gadget} \mbox{N-body} simulation from the same initial conditions, consisting of $2048^3$ dark matter particles, 8 times fewer than CoDa I, but with a force resolution that is around 20 times better. CoDa I-DM2048 contains about 5 times as many $\simeq 10^{7.5} \text{ M}_{\odot}$ halos per unit mass per comoving volume as CoDa I. This may be an effect of CoDa's force resolution limitations, as the discrepancy is smaller for higher mass halos (e.g. a factor of 2 at $10^8\text{  M}_{\odot}$, and only a 10\% difference at $10^{8.5}\text{  M}_{\odot}$). Therefore, in order to minimize any errors that could result from this, we restrict our analysis to halos with $M>10^{8} \text{  M}_{\odot}$. This halo mass range accounts for almost all of the star-formation in the simulation, however, so this does not limit our analysis in any significant way. For further discussion of the CoDa I and CoDa I-DM2048 halo mass functions, see the appendix in \citet[]{Ocvirk}.

\subsection{Results}

\begin{figure*}
    \centering
    \includegraphics[width=0.59\textwidth]{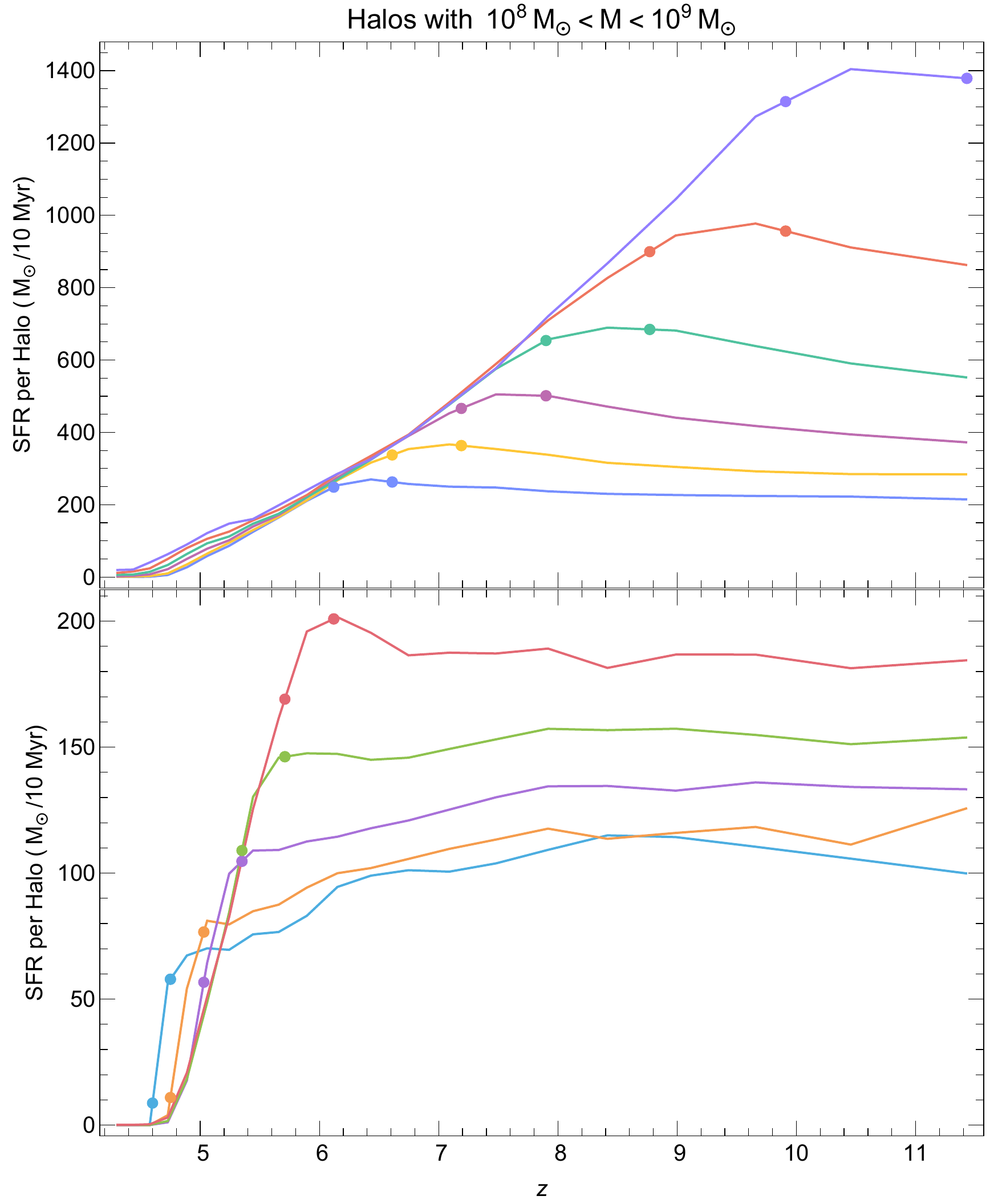}
    \caption{Average star formation rates per halo vs redshift for low-mass ($10^{8-9} \text{ M}_\odot$) halos, with different curves representing different bins of reionization redshift, $z_{re}$. The two dots on each curve indicate the range of $z_{re}$ for each bin. For example, the top blue curve in the top panel includes halos with reionization redshifts in the range $9.91<z_{re}<11.43$, and so the points on the curve are located at $z=9.91$ and $z=11.43$. Results are plotted in two panels so that all curves can be clearly distinguished on a linear plot.}
    \label{fig:low}
\end{figure*}
\begin{figure}
    \centering
    \includegraphics[width=\columnwidth]{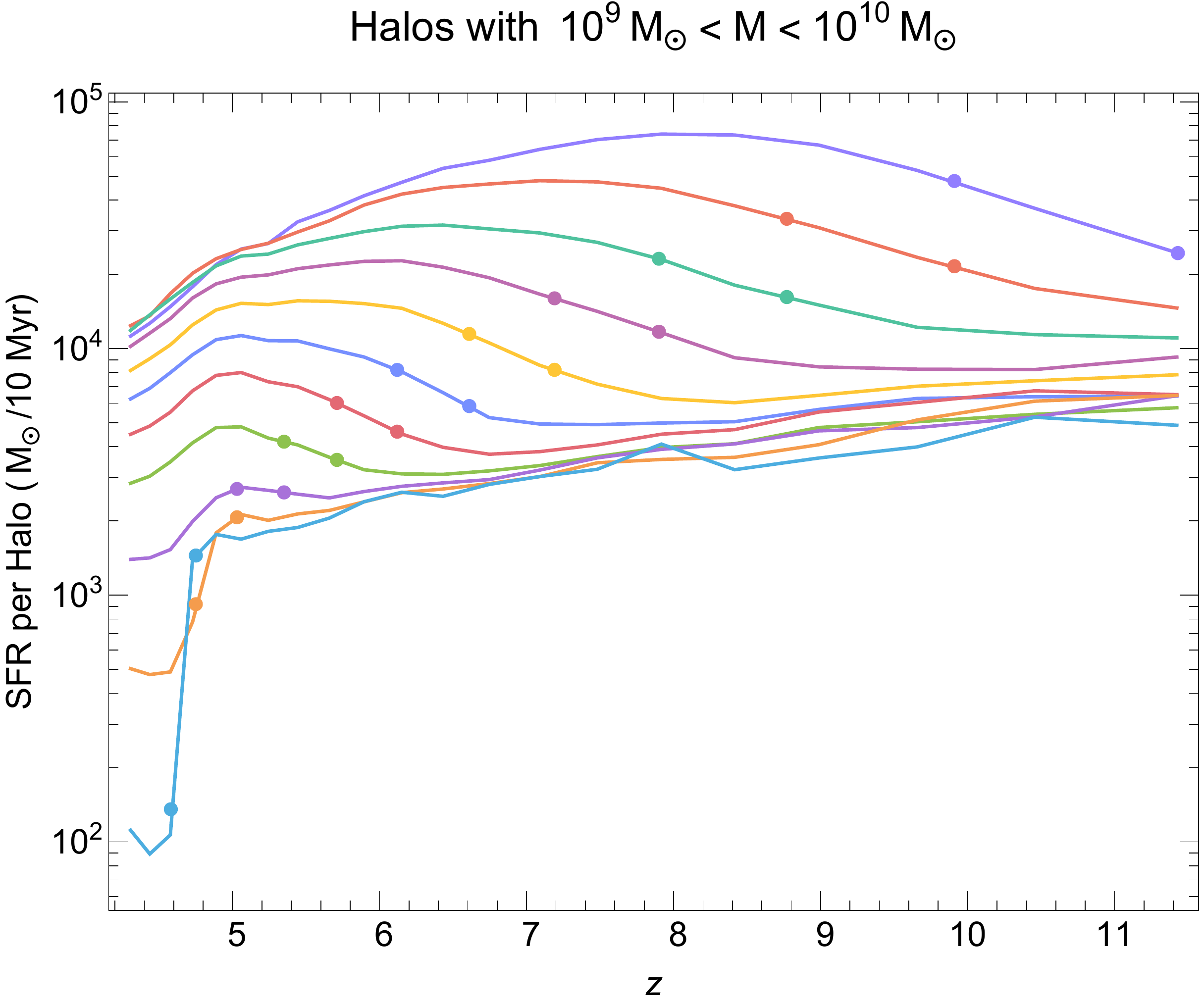}
    \caption{Same as Fig.~\ref{fig:low}, but for intermediate-mass ($10^{9-10} \text{ M}_\odot$) halos.}
    \label{fig:intermediate}
\end{figure}
\begin{figure}
    \centering
    \includegraphics[width=\columnwidth]{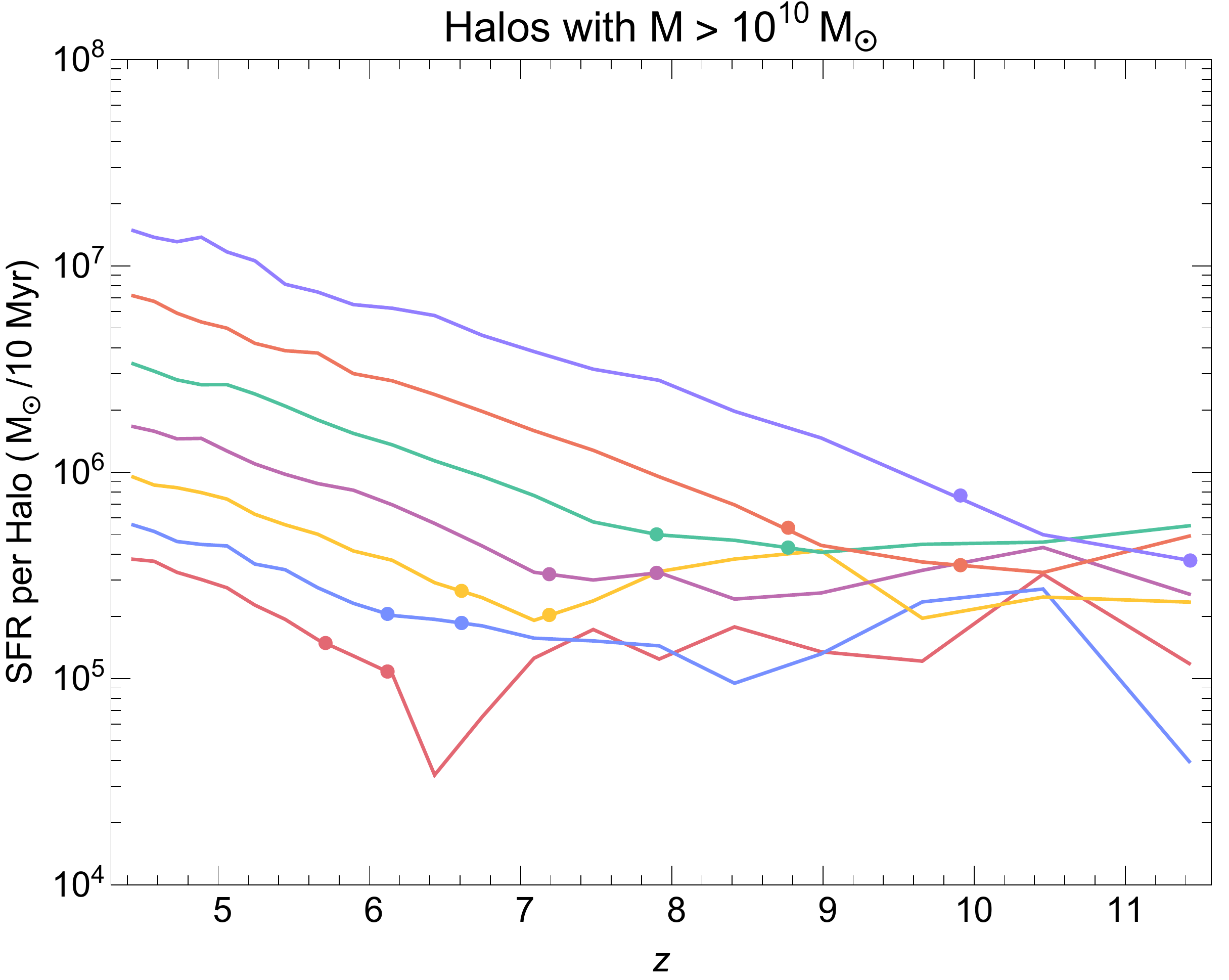}
    \caption{Same as Fig.~\ref{fig:intermediate}, but for high-mass ($10^{10+} \text{ M}_\odot$) halos. Some late $z_{re}$-bins are omitted because they are too noisy; high-mass halos are rare in regions that reionized late (voids).}
    \label{fig:high}
\end{figure}

\subsubsection{Local Reionization and the Suppression of Star Formation}
We analyzed the CoDa I simulation results at 103 redshifts, ranging from $z=11.43$ (at which the globally-averaged $X_\text{ion}=1.8\times10^{-3}$) to $z=4.23$, equally-spaced in time intervals of 10 Myrs.
At each of these time-steps, we grouped halos into reionization redshift bins and calculated their instantaneous SFR. This data is shown for halos in three mass ranges: $10^{8-9} \text{  M}_{\odot}$ (Fig.~\ref{fig:low}), $10^{9-10} \text{ M}_{\odot}$ (Fig.~\ref{fig:intermediate}), and $10^{10+} \text{ M}_{\odot}$ (Fig.~\ref{fig:high}). Each reionization redshift bin is plotted on a separate curve, and the $z_{re}$ interval encompassed by that bin is bounded by two dots on the curve. For example, the top purple curve in the figures includes halos with reionization redshifts in the range $9.91<z_{re}<11.43$, so the dots on the curve are located at $z=9.91$ and $z=11.43$. The reionization redshifts of all the halos represented by that curve are between these two dots. 

The results in Figs.~\ref{fig:low} -~\ref{fig:high} show that the SFRs in low-mass halos are reduced by reionization feedback while those in high-mass halos are not. 
The suppression of low-mass halos ($10^{8}\text{ M}_\odot<M<10^{9}\text{ M}_\odot$) in Fig.~\ref{fig:low} clearly {\it follows their local reionization}: for each $z_{re}$ bin, the SFR per halo is steady (i.e. roughly constant) until $z=z_{re}$, from which point it drops sharply. This suppression is caused by reionization feedback. Fig. 8 in \citet[]{Ocvirk} indicates that it cannot be the result of supernova (SN) feedback alone, since the comparison there of two simulations, one with both RT and SN feedback and one with only SN feedback but no RT, shows that suppression of SFR in low-mass halos is overwhelmingly stronger when RT is included. Moreover, SN feedback is already operating inside the low-mass halos in our Fig.~\ref{fig:low} {\it prior} to their neighborhood's reionization, when the SFR per halo is roughly constant. What distinguishes this prior phase from that {\it after} reionization, when the SFR drops sharply, is the onset of radiative feedback. 

As expected, high-mass halos ($M>10^{10}\text{ M}_\odot$) in Fig.~\ref{fig:high}, by contrast, go through local reionization without experiencing any suppression at all. Instead, the SFR per halo increases with time in this high-mass range. 
Intermediate-mass halos ($10^{9}\text{ M}_\odot<M<10^{10}\text{ M}_\odot$) in Fig.~\ref{fig:intermediate} see a modest increase in their average SFR through their local reionization time, followed by a delayed, modest decline. In the regions that reionize last, however, the decline, even for this intermediate-mass range, is sharp, and there is no appreciable increase in the SFR prior to that decline. Indeed, the intermediate-mass-halo SFR curves for these late-reionizing regions resemble the low-mass-halo SFR curves.

In all three mass bins, regions that reionize earlier have higher SFRs than regions that reionize later, though this is true only prior to local reionization in the low-mass bin. After their local reionization, low-mass halos in all $z_{re}$ bins seem to have roughly the same average SFR if compared at the same redshift. There appears to be a universal curve along which suppressed-halo SFRs lie once they begin their decline, regardless of the halo's reionization redshift, despite the fact that the SFR varies substantially across different $z_{re}$ bins prior to suppression.
In particular, each pre-reionization SFR level for a given $z_{re}$ bin meets the universal curve at $z\approx z_{re}$ and, thereafter, follows the universal curve towards lower $z$.

\begin{figure}
    \centering
    \includegraphics[width=\columnwidth]{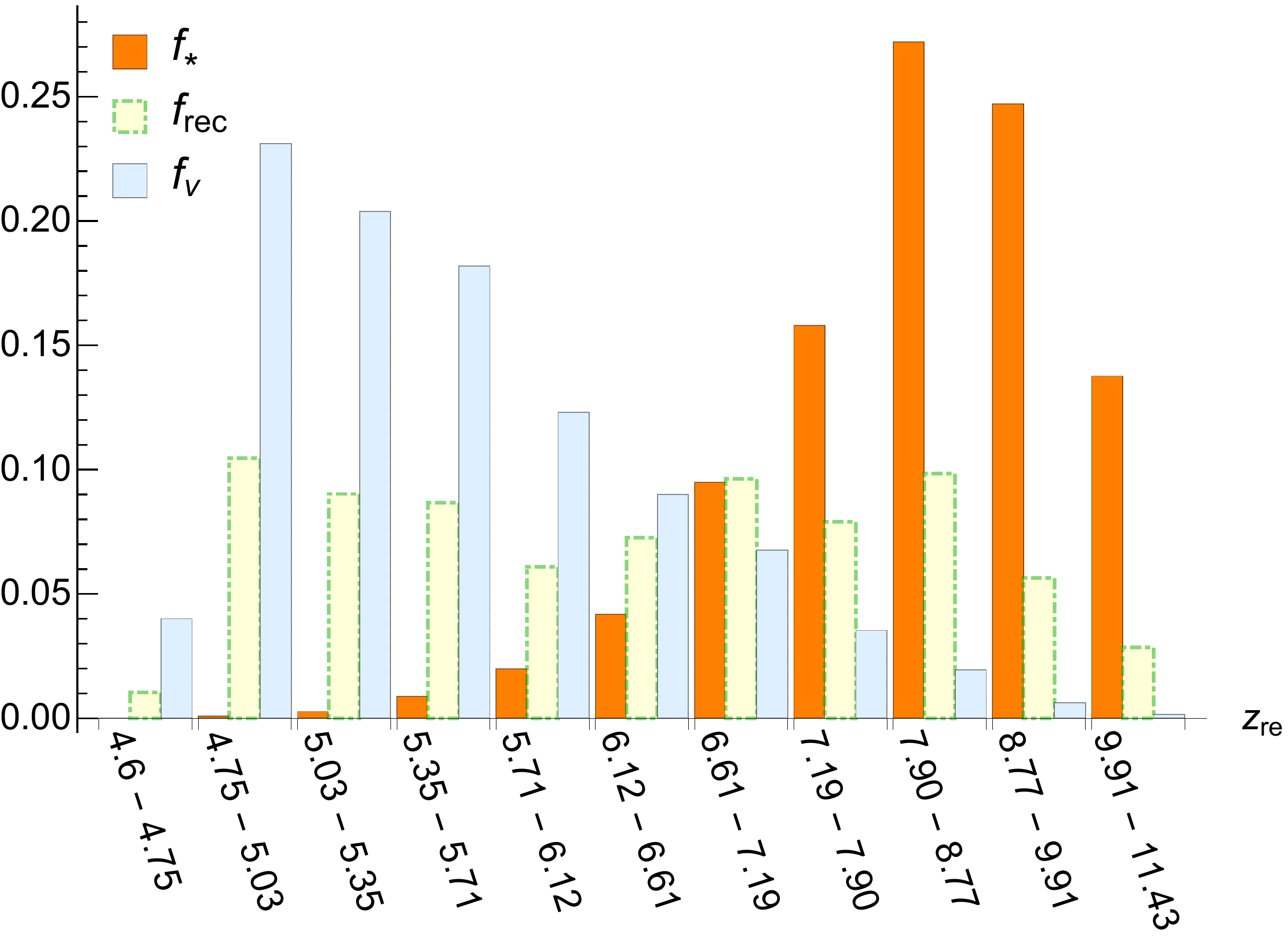}
    \caption{Fractions of the total stellar mass, $f_\star$, the total number of recombinations, $f_\text{rec}$, and the total volume, $f_\text{v}$, associated with the cells in each reionization redshift ($z_{re}$) bin.}
    \label{fig:starmassfrac}
\end{figure}

Since early-reionizing regions are rare (see Fig.~\ref{fig:zre}) and the first to have their SFR suppressed, one might expect that they contribute very little to the global production of stars. But, in fact, we find that the opposite is true. For instance, the first 6.25\% of cells to reionize (those with $z_{re}>7.19$, represented by the top 4 curves in Fig.~\ref{fig:low},~\ref{fig:intermediate}, and~\ref{fig:high}) produced 83.1\% of the total stellar mass in the simulation. On the other hand, the last 87\% of cells to reionize (those with $z_{re}<6.61$, represented by the bottom 6 curves in Fig.~\ref{fig:low} and~\ref{fig:intermediate}) produced just 7.4\% of the total stellar mass. The fraction of the total stellar mass produced in each $z_{re}$ bin, $f_\star$, is shown in Fig.~\ref{fig:starmassfrac}, along with the fraction of the total volume encompassed by all the cells in that $z_{re}$ bin, $f_\text{v}$. The rare, early-reionizing regions clearly dominate the total production of stars. This is due to the fact that these regions are more overdense and produce many more halos and relatively more high-mass halos (both early on and over time) than regions that reionize later, as we will show in the next section.

The rate of production of ionizing photons is proportional to the SFR, while the total number of ionizing photons produced over time is proportional to the total stellar mass. It might be tempting to conclude from the results above, therefore, that early-reionizing regions are the dominant contributors to the ionizing UV background responsible for global reionization. However, this neglects the possibility of local consumption of ionizing photons by their absorption. To account for this effect, we can count the recombinations in each computational cell that reflect the cancellation of photoionizations there. 
We show the fraction of recombinations that occurred in all the computational cells encompassed by each $z_{re}$ bin, $f_\text{rec}$, in Fig.~\ref{fig:starmassfrac}, for which the Case B recombination rate of hydrogen was calculated as
\begin{equation}
    \mathcal{R}=\alpha(T)n_en_{p}=\alpha(T)(n_p)^2,
    \label{recombrate}
\end{equation}
where $\alpha(T)=2.6\times10^{-13}(T/10^4 \text{ K})^{-0.7}\rm{ s^{-1} cm^3}$ for a gas temperature $T$, electron number density $n_e$, and ionized hydrogen number density $n_{p}=X_\text{ion}n_\textsc{h}$ \citep[][]{Spitzer}.

Consider the fractions $f_\star$ and $f_\text{rec}$ in a given region. Assume that most of the ionizing photons that were produced during the epoch of reionization were absorbed by the time reionization ended (or shortly thereafter).
On average, per unit volume, the number of ionizing photons required to finish reionization was equal to the number of H atoms {\it plus} the total number of recombinations.
In that case, photons produced in one region may either have been absorbed in the same region or in another region.
In Appendix~\ref{appendix}, we show that regions in which $f_\star<f_\text{rec}$ could not have produced enough ionizing photons to finish and maintain their own local reionization. They must have `imported' ionizing radiation from external sources in order to finish and/or maintain their own reionization. Conversely, as also shown in Appendix~\ref{appendix}, regions in which $f_\star\gg f_\text{rec}$ must have produced a `surplus' of ionizing radiation (i.e. more than enough to complete and maintain their local reionization). They must be net `exporters' of ionizing photons to other regions, therefore. According to Fig.~\ref{fig:starmassfrac}, late-reionizing regions are the ones in which $f_\star<f_\text{rec}$, so they must be net `importers', while early-reionizing regions are generally net `exporters'. These results validate the conclusion that early-reionizing regions are the dominant contributors of the UV photons that reionized the universe.

\subsubsection{Local Reionization and its Dependence on the Local Halo Mass Function and Overdensity}
\label{HMFandOD}

\begin{figure*}
    \centering
    \includegraphics[width=0.70\textwidth]{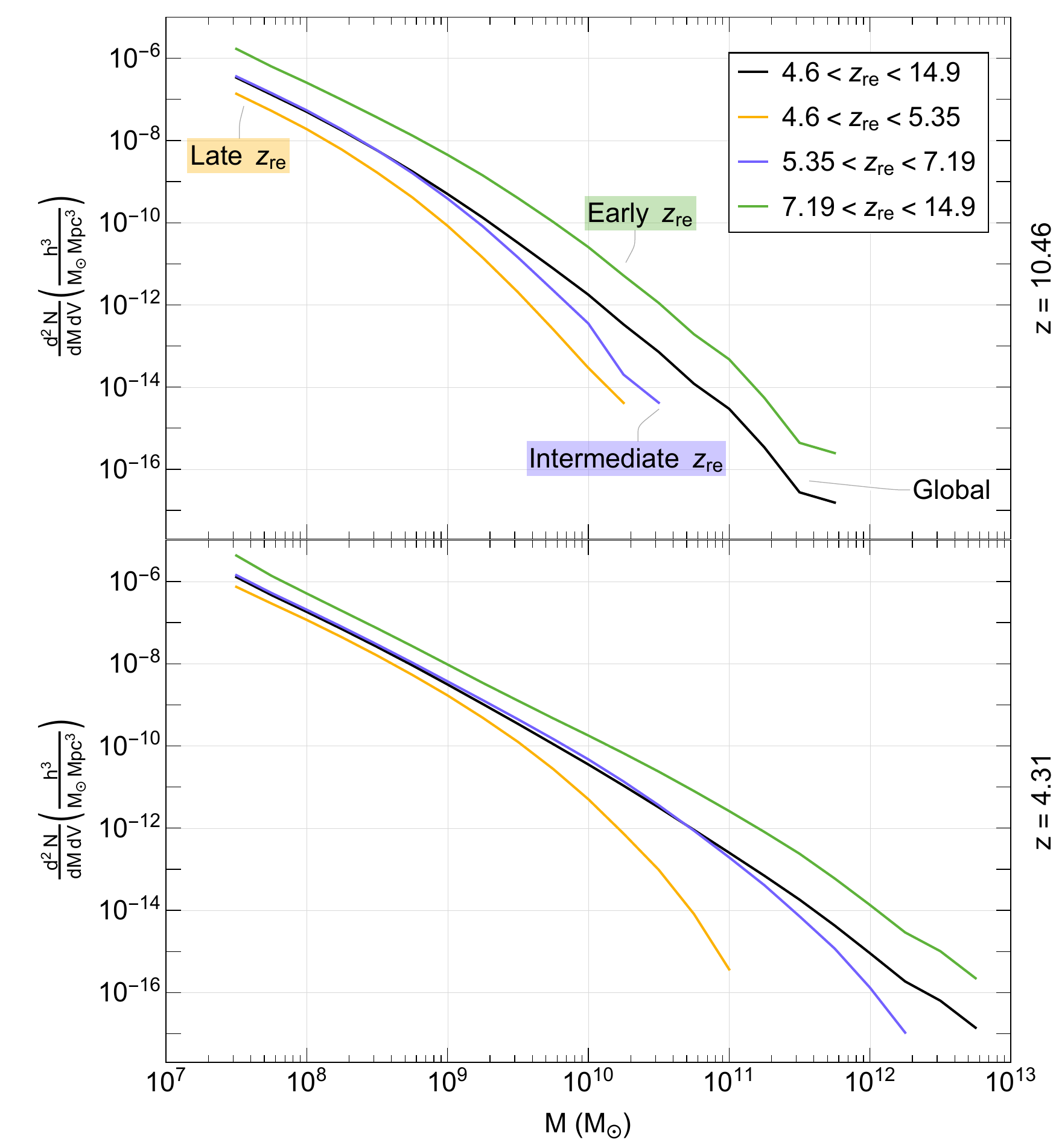}
    \caption{Local halo mass functions for halos grouped into each of three broad reionization redshift bins, plotted at two redshifts: 10.46 (top) and 4.31 (bottom). The global halo mass function is also plotted (black). Halo mass is binned with a bin size of $\Delta[\rm{log}(M/M_\odot)]=0.25$.}
    \label{fig:HMF}
\end{figure*}

\begin{figure}
    \centering
    \includegraphics[width=\columnwidth]{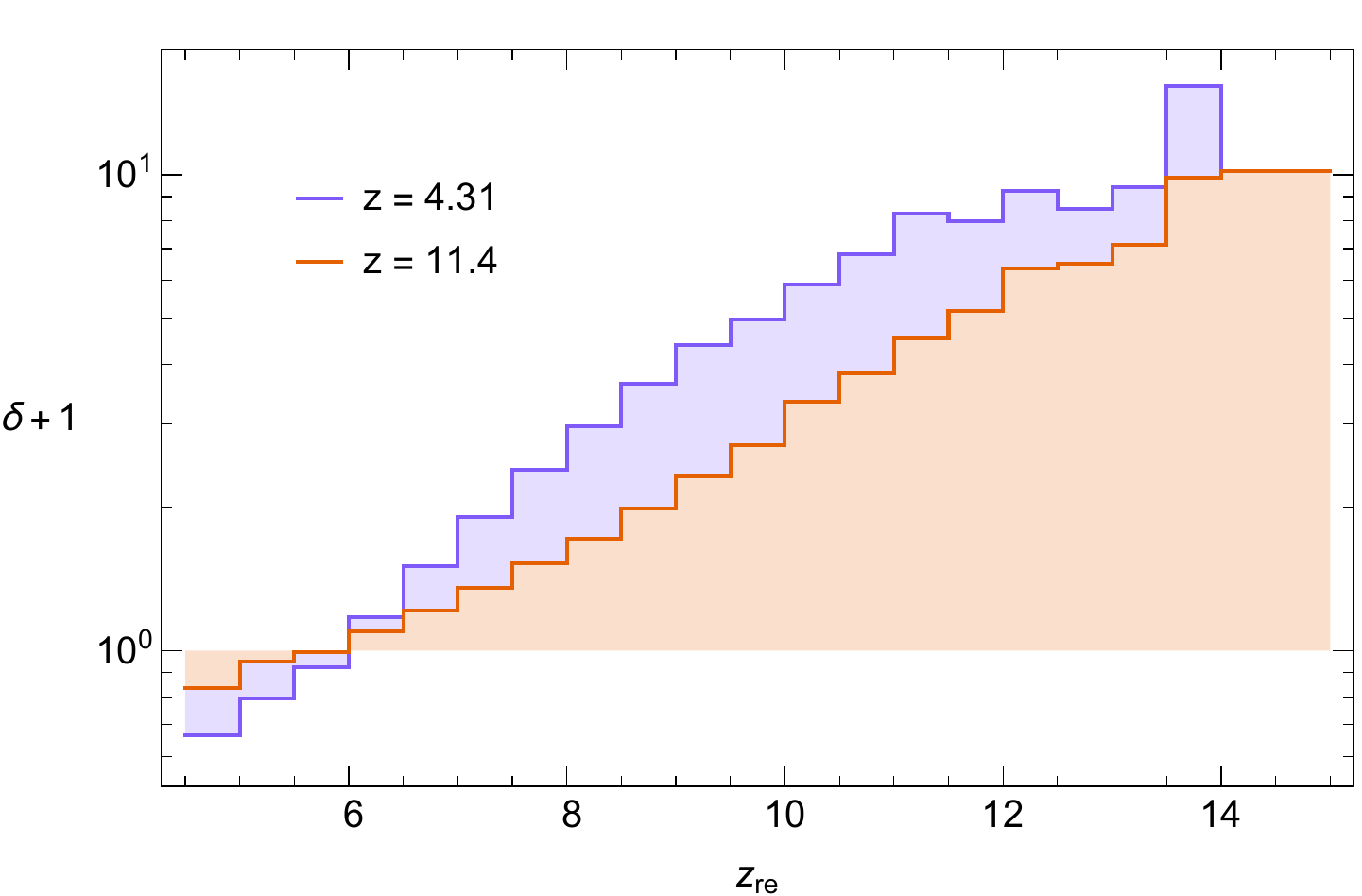}
    \caption{The relationship between local mass overdensity, $\delta$, and local reionization redshift, $z_{re}$, at early ($z=11.4$) and late ($z=4.31$) times. Overdensity $\delta \equiv \rho/\bar{\rho}-1$ where $\rho$ is the mass density of the region, and $\bar{\rho}$ is the average mass density of the universe. This demonstrates that the earliest regions to reionize correspond to density peaks, while the latest correspond to voids.}
    \label{fig:ODvZre}
\end{figure}
\begin{figure}
    \centering
    \includegraphics[width=\columnwidth]{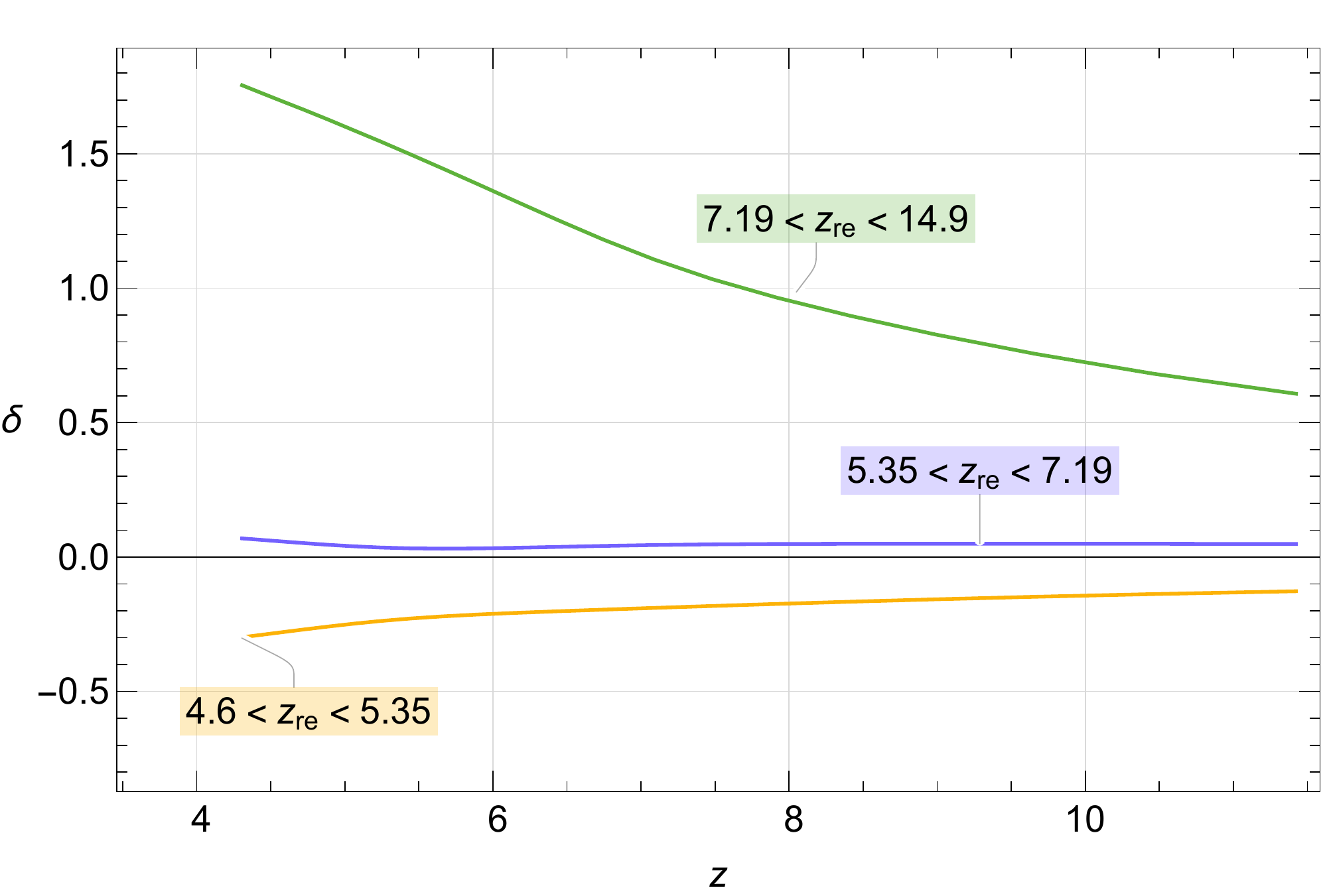}
    \caption{Local mass overdensity, $\delta$, grouped into three broad reionization redshift bins, and plotted against redshift to illustrate the `rich-get-richer' scheme of cosmological structure formation. Initially overdense, early-reionizing regions continue to get more overdense, while initially underdense, late-reionizing regions continue to get more underdense.}
    \label{fig:od}
\end{figure}

We have shown that suppression of the SFR per halo in low-mass halos always {\it follows} the reionization of their local IGM. In the process, however, we have also learned that there are interesting differences between regions that reionize earlier or later, even in their SFRs {\it prior} to their local reionization, and even for halos too massive to experience suppression.
Our grouping of cells into reionization redshift bins allows us to study these differences. One important difference is their local halo mass function (HMF), the number of halos per unit halo mass per comoving volume, shown at an early time and a late time in Fig.~\ref{fig:HMF}. At all redshifts, the HMF for regions that reionized early are higher (corresponding to a greater density of halos) and have a larger turn-over mass (corresponding to a higher fraction of massive halos) than regions that reionized later.
Also shown for comparison in Fig.~\ref{fig:HMF} is the globally-averaged halo mass function. According to Fig.~\ref{fig:HMF}, the mass function in earlier-reionizing regions is higher than the globally-averaged mass function, while in the later-reionizing regions the mass function is always below the global one (and with a lower-mass turn-over). Over time these contrasts between the local and global mass functions grow larger. 

The correlation of local reionization redshift with local HMF in Fig.~\ref{fig:HMF} reflects the fact that the production rate of stars (and, hence, of ionizing photons) is higher in regions with a higher HMF, so reionization ends earlier there. Since a higher local HMF means a higher structure formation rate, and the latter is higher if the local overdensity is higher, we expect local reionization to be earlier where overdensity is higher.
This is indeed the case, as can be seen in Fig.~\ref{fig:ODvZre}. 
The correlation between overdensity and reionization time was noted before by \citet[]{Iliev06}, based upon large-scale N-body + RT simulations of reionization.
As shown in Fig.~\ref{fig:ODvZre}, the {\it earliest} patches to reionize are the density peaks, while the {\it latest} are in the voids. As shown in Fig.~\ref{fig:od}, they follow a `rich-get-richer, poor-get-poorer' scheme: as time goes on, early-reionizing patches become increasingly overdense, while late-reionizing patches become increasingly underdense. Structure formation in voids never catches up to that of the density peaks. 

The shape of the HMF explains many features of the SFR plots in Figs.~\ref{fig:low} -~\ref{fig:high}. For all three mass ranges, the SFR curves are stacked (prior to suppression in the low-mass range) in order of their $z_{re}$ bins, with the earliest $z_{re}$ bin at the top (i.e. greatest SFR). This is partially due to the fact that the early $z_{re}$ bins have more halos with masses close to the upper bound of each of the mass ranges, as can be seen in the HMF. Since the SFR (per halo) follows an $M^{5/3}$ power law prior to suppression, the relative abundance of more massive halos in early-reionized regions provides a boost to the SFR there. Conversely, the latest $z_{re}$ bins have far fewer massive halos, which is why they do not show a delayed suppression in the intermediate-mass range curves in Fig.~\ref{fig:intermediate} -- the late-$z_{re}$ curves there resemble the SFR curves in the low-mass range plot because they have comparatively few halos that are significantly more massive than the upper bound of the low-mass range. The scarcity of more massive halos in these low-$z_{re}$ bins, especially at early redshifts, made them too noisy to include in Fig.~\ref{fig:high}.

\begin{figure}
    \centering
    \includegraphics[width=\columnwidth]{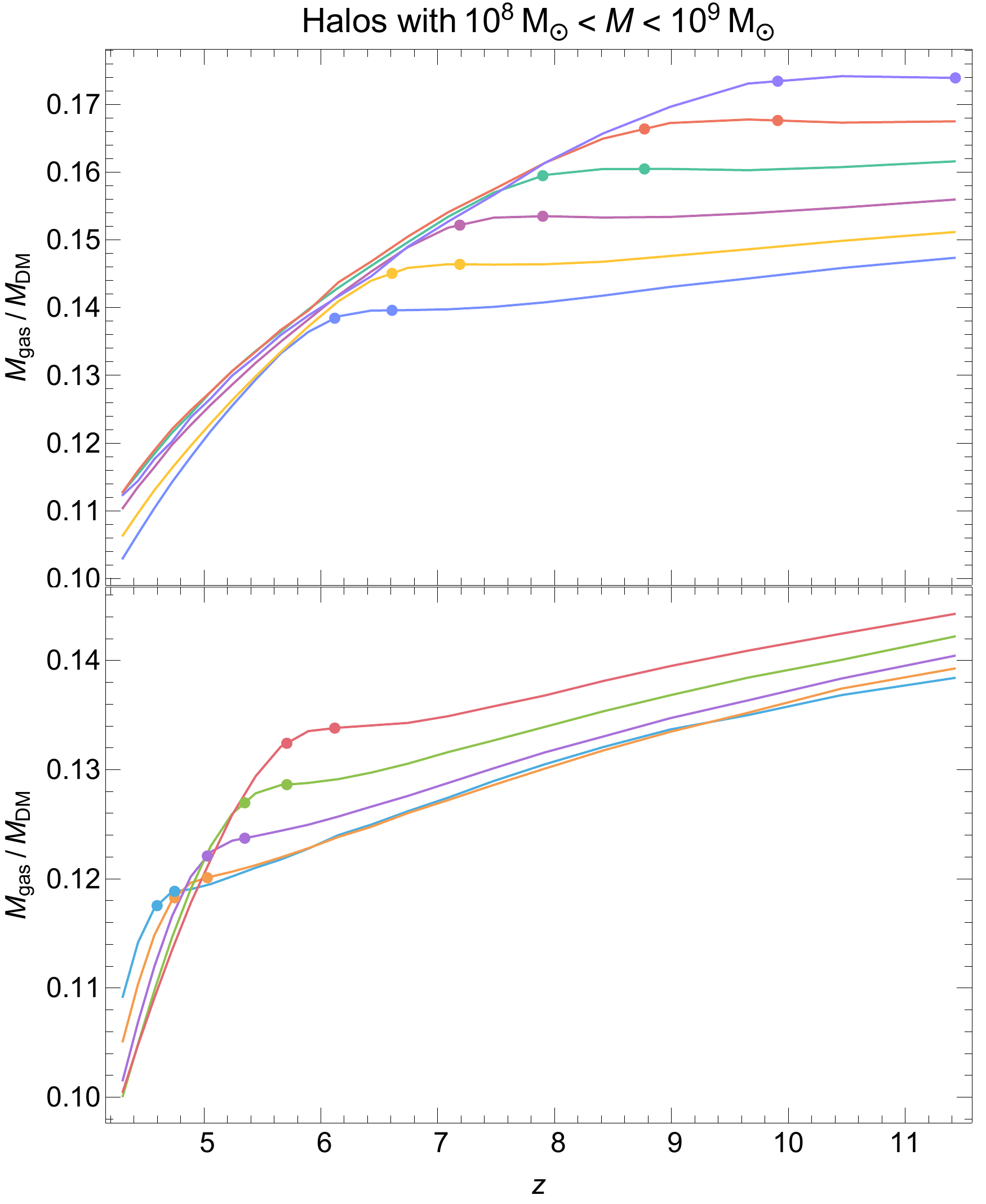}
    \caption{Average halo gas-to-dark matter ratio vs redshift for low-mass ($10^{8-9} \rm{ M_\odot}$) halos, with different curves representing different $z_{re}$ bins. The two dots on each curve indicate the range of $z_{re}$ for each bin, in the same way as Fig.~\ref{fig:low}. The cosmic mean value is $\langle M_\text{gas}/M_\textsc{dm} \rangle \simeq \Omega_\text{bary}/\Omega_\textsc{dm} \simeq 0.2$ (since $\langle M_\star/M_\text{gas}\rangle \ll 1$).}
    \label{fig:lowgf}
\end{figure}
\begin{figure}
    \centering
    \includegraphics[width=\columnwidth]{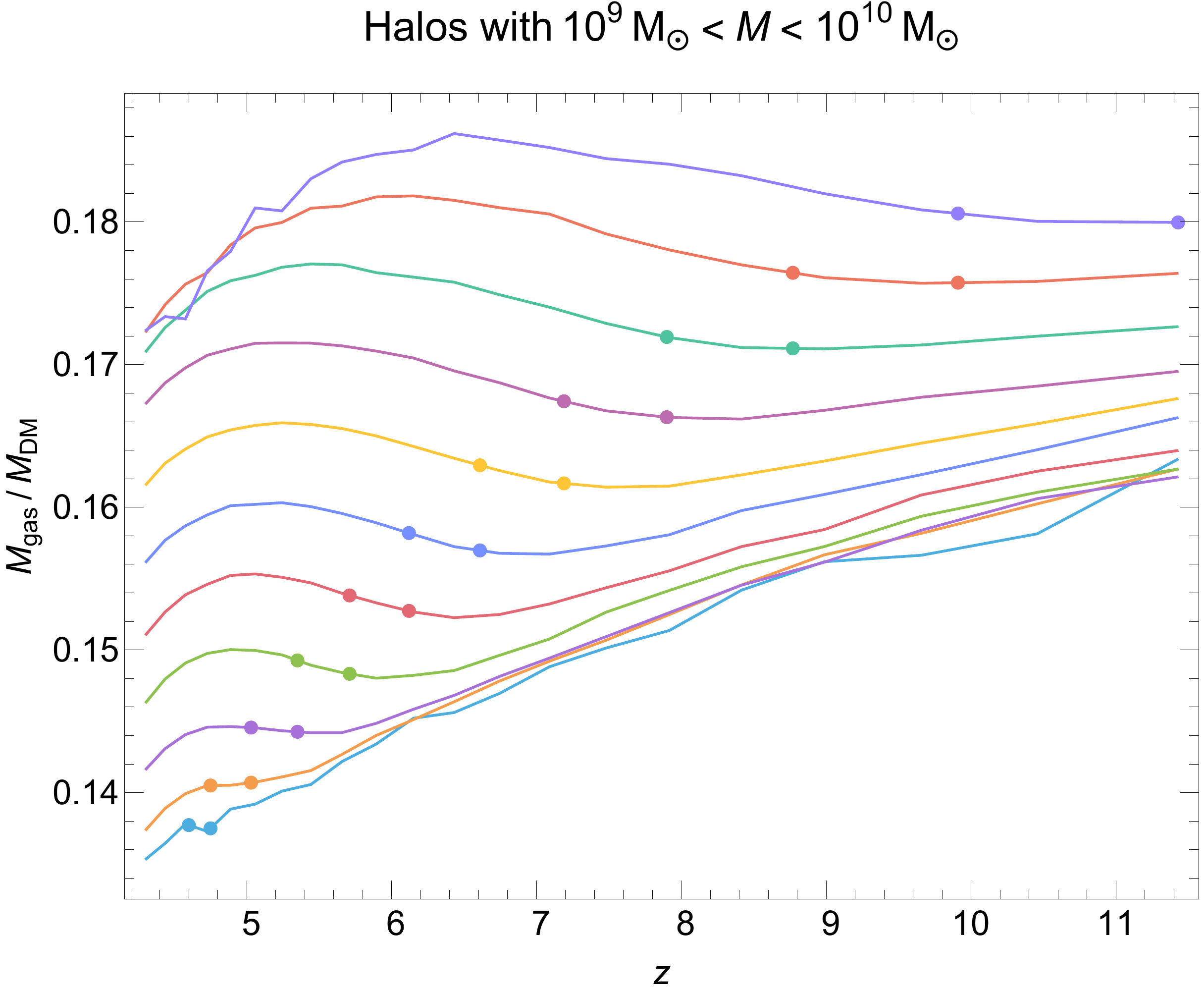}
    \caption{Same as Fig.~\ref{fig:lowgf}, but for intermediate-mass ($10^{9-10} \rm{ M_\odot}$) halos.}
    \label{fig:intermediategf}
\end{figure}
\begin{figure}
    \centering
    \includegraphics[width=\columnwidth]{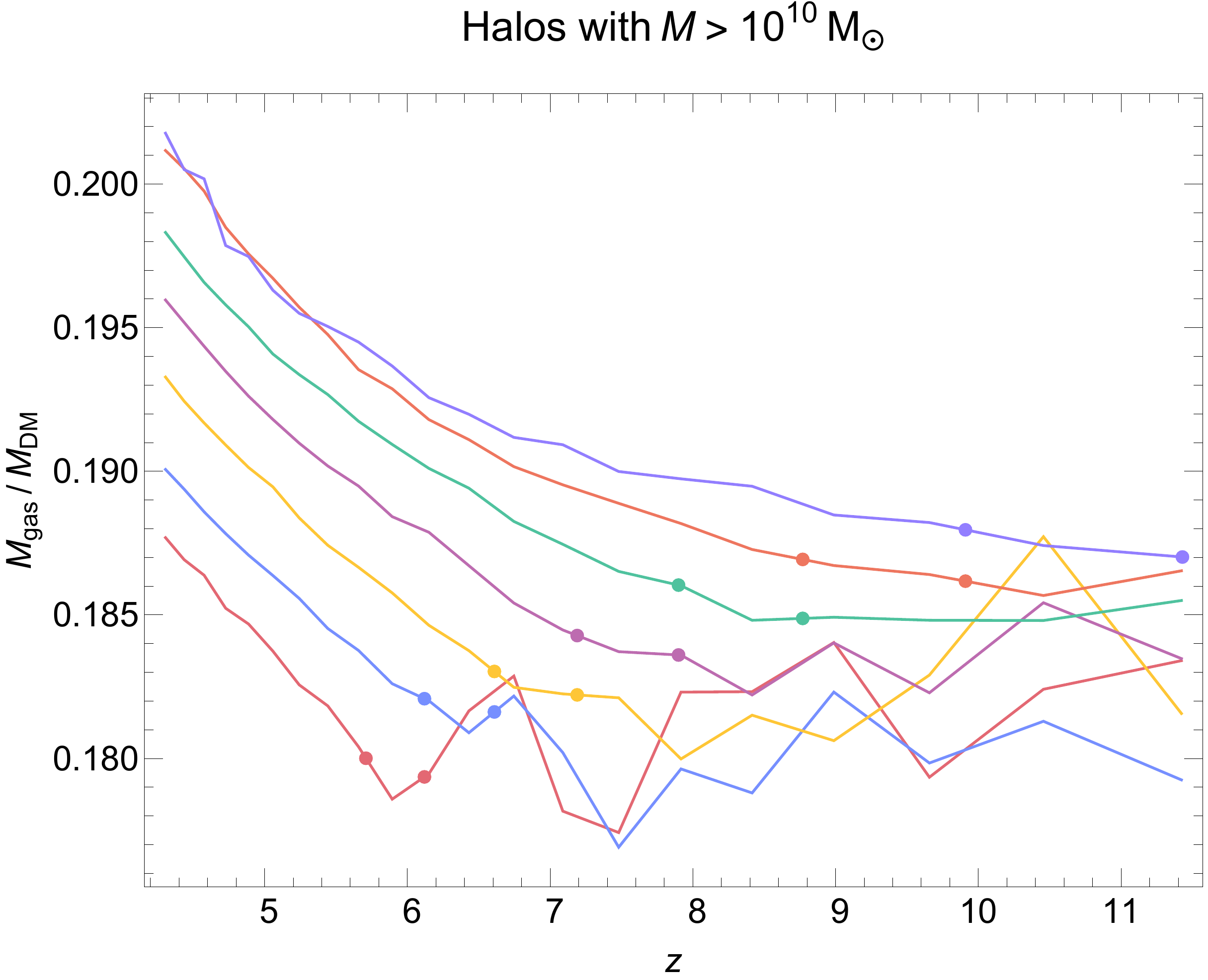}
    \caption{Same as Fig.~\ref{fig:lowgf}, but for high-mass ($10^{10+} \rm{ M_\odot}$) halos. Some late $z_{re}$ bins are omitted because they are too noisy.}
    \label{fig:highgf}
\end{figure}

\subsubsection{Local Reionization and the Halo Gas-to-Dark Matter Ratio}

The effect of local reionization on low-mass halos can also be observed via the ratio of mass in gas to mass in dark matter in a halo, $M_\text{gas}/M_\textsc{dm}$. Since the loss of gas due to outflow and suppressed infall is what is thought to cause SFR suppression, $M_\text{gas}/M_\textsc{dm}$ should reflect a causal relationship between local reionization and suppression, as well. 
Since the mass in stars is always much smaller than that in gas, $M_\text{gas}/M_\textsc{dm} \simeq M_\text{bary}/M_\textsc{dm} = f_\text{bary}\Omega_\text{bary}/\Omega_\textsc{dm}$, where $f_\text{bary}$ is the fraction of the cosmic mean baryon-to-dark matter ratio. Since $\Omega_\text{bary}/\Omega_\textsc{dm}=0.197$ for CoDa I parameters, this means $f_\text{bary}\simeq5M_\text{gas}/M_\textsc{dm}$.
Roughly speaking, if \mbox{$f_\text{bary}\simeq1=$ constant}, at all times for a given halo, then its baryonic gas content was not suppressed. However, since we do not here analyze the Lagrangian mass assembly history of individual halos, we must diagnose the effects of suppression by tracking the average value of $M_\text{gas}/M_\textsc{dm}$ of all the halos identified at each time, grouped according to their reionization redshift, just as we did for the instantaneous SFR per halo, and plot this quantity versus $z$, instead. 

We calculated $M_\text{gas}/M_\textsc{dm}$ for each halo as follows. We constructed a dark matter density grid by smoothing the N-body particle data onto the same slightly-coarsened grid of $2048^3$ cells as we created to smooth the gas density from the original $4096^3$ simulation cells\footnote{
We note that, if instead of using the dark matter density grid, we had calculated $M_\textsc{dm}$ simply by adding together the masses of all the N-body particles the FOF algorithm identified as belonging to a halo, the resulting gas-to-dark matter ratios would not have been reliable. This is because the volumes occupied by these halo particles need not be the same as that of the gas cells inside the halo's $R_{200}$ sphere. In that case, $M_{\text{gas}}/M_{\textsc{dm}}$ typically exceeds the cosmic mean value ($\simeq 0.197$), which seems unphysical and, therefore, inaccurate. By using the cell-wise data for {\it both} the dark matter  and gas density fields, instead, we ensure that the calculation of $M_{\textsc{dm}}$ considers the halo to occupy the same volume as the calculation of $M_{\text{gas}}$.}.
For a given halo $i$, we identify the $n(i)$ cells within the spherical volume defined by its halo $R_{200}$ radius\footnote{Of course, a spherical halo will not encompass an integer number of cubic cells. In the gas fraction calculation, we chose to include all cells that are partially encompassed by the $R_{200}$ sphere as well, such that the total volume of the `halo' is somewhat greater than $4\pi R_{200}^3/3$.}
and sum the gas and dark matter masses, respectively, of those cells:
\begin{equation}
    \Big(\frac{M_\text{gas}}{M_\textsc{dm}}\Big)_i =\frac{\sum_{j=1}^{n(i)}\rho_{\rm{gas}} (i,j)V_\text{cell}}{\sum_{j=1}^{n(i)}\rho_{\textsc{dm}} (i,j)V_\text{cell}} = \frac{\sum_{j=1}^{n(i)}\rho_{\rm{gas}} (i,j)}{\sum_{j=1}^{n(i)}\rho_{\textsc{dm}} (i,j)},
\end{equation}
where $\rho_{\rm{gas}}(i,j)$ and $\rho_{\textsc{dm}}(i,j)$ are the gas and dark matter densities in the $j^\mathrm{th}$ cell of the $i^\mathrm{th}$ halo, and $V_\text{cell}$ is the volume of a cell.
Then, for each $z_{re}$ bin containing $N(z_{re})$ halos,
we define the instantaneous, $z_{re}$-binned, average halo gas-to-dark matter ratio as:
\begin{equation}
    \frac{M_\text{gas}}{M_\textsc{dm}} =\frac{1}{N(z_{re})}\sum_{i=1}^{N(z_{re})}\Big(\frac{M_\text{gas}}{M_\textsc{dm}}\Big)_i.   
\end{equation}

We plot this $M_\text{gas}/M_\textsc{dm}$ vs redshift for each of the $z_{re}$ bins in Figs.~\ref{fig:lowgf} -~\ref{fig:highgf}, for the same halo mass bins as shown in the SFR plots in Figs.~\ref{fig:low} -~\ref{fig:high}. The $M_\text{gas}/M_\textsc{dm}$ curves in Figs.~\ref{fig:lowgf} -~\ref{fig:highgf} follow the same pattern as do those of the SFRs per halo in Figs.~\ref{fig:low} -~\ref{fig:high}. For halos below $10^9 \text{ M}_\odot$, the gas-to-dark matter ratio is relatively steady prior to the local reionization redshift but plunges sharply once local reionization occurs. The pre-reionization `steady' value actually declines a bit during the pre-reionization phase for the late-reionizing regions, but then declines much more abruptly for these regions following their local reionization. For the intermediate-mass halos with $10^9\text{ M}_\odot<M<10^{10}\text{ M}_\odot$, there is a similar downward tilt to the pre-reionization curves for the regions with later-reionization redshifts, but the ratios actually increase somewhat for the halos in most of the \mbox{$z_{re}$-bins}, after their reionization, to a peak, followed by a delayed turn-down. For the high-mass halos with $M>10^{10}\text{ M}_\odot$, however, $M_\text{gas}/M_\textsc{dm}$ grows continuously following $z_{re}$. For halos in every mass bin, the gas-to-dark matter ratios prior to local reionization are higher in regions that reionized earlier.

These results are consistent with the expectations of suppression in low-mass halos of baryonic infall relative to that of the dark matter, either by Jeans-mass filtering of the baryonic perturbations in the IGM, or suppressed accretion from the IGM onto a dark matter halo, or of feedback-induced outflows from the halos. The suppressed gas fraction in low-mass halos following their local reionization is also physically consistent with the SFR suppression results reported above, including the fact that the gas fractions and SFRs are both higher for halos in regions that reionized earlier, except for low-mass halos during their post-reionization phase.

\section{Summary and Discussion}
\label{conclusion}
We have analyzed the results of the CoDa I fully-coupled radiation-hydrodynamical simulation of reionization and galaxy formation to demonstrate the causal relationship between reionization feedback and the suppression of star formation in low-mass halos during the EOR. This required us to find the dependence of the star formation rates per halo of different halo masses on the timing of reionization in their local neighborhood. For this purpose we created a reionization redshift field based on the redshift at which the gas in each computational cell reached an ionized fraction of 0.9. We smoothed this field to a grid coarse enough to sample the reionization redshift of the IGM surrounding each halo, but fine enough to track the impact of the inhomogeneous and asynchronous nature of reionization on individual halos. This gave us the local reionization redshift for each halo with which we could quantify the difference between halo properties before and after their local reionization.
We show that star formation rates and gas-to-dark matter ratios in low-mass ($10^{8-9} \text{ M}_\odot$) galaxies was steady prior to their local reionization, following which they sharply declined. This effect is not seen in high-mass ($10^{10+} \text{ M}_\odot$) galaxies, but a moderate and delayed suppression is seen in the intermediate-mass range ($10^{9-10} \text{ M}_\odot$), except in the last regions to reionize, where the suppression is sharper. 

These results are in accordance with the notion that baryonic effects during the EOR may explain the `missing satellites' problem posed by $\Lambda$CDM N-body simulations, by suppressing star formation in low-mass halos after their neighborhood was reionized, thereby reducing their detectability. These results also support the assumption made in large-scale N-body + RT simulations of inhomogeneous reionization that halos with $M<10^9\text{ M}_\odot$ were suppressed if located inside an ionized patch of the IGM \citep{Iliev07,Iliev12,iliev2014,Ahn12,Dixon16}. 

Our ability to distinguish the progress and impact of reionization {\it locally} here has enabled us, not only to demonstrate the causal connection between reionization and low-mass halo suppression, but also to show that the time at which local reionization takes place is correlated with local matter overdensity and with the corresponding local variations of the halo mass function. In addition, we found that, with the exception of suppressed low-mass halos, the SFRs and gas-to-dark matter ratios of halos of any given mass are higher in regions that reionized earlier. For the low-mass halos, after their local reionization, when suppression takes place, curves of SFR and $M_\text{gas}/M_\textsc{dm}$ per halo vs $z$ thereafter decline by following a universal curve which is independent of $z_{re}$.

We also found that relatively rare, highly overdense and, consequently, highly developed regions, which were the first to reionize, produced the vast majority of stars during the EOR, while not consuming more ionizing radiation than regions that reionized later. These early-reionizing regions must have exported a surplus of ionizing radiation, therefore, which helped to reionize and/or maintain the reionization of the late-reionizing regions.
Future observations of high-redshift galaxies which look for the 
inhomogeneous distribution of suppressed and gas-poor 
low-mass dwarf galaxies and partially-suppressed dwarf
galaxies of intermediate mass predicted here will be an excellent
probe of the EOR.

Since the CoDa I simulation used a constrained realization of the initial conditions that represents the Local Universe, we can also apply the results found here to predict relic signatures of suppression in the present day galaxies of the Local Universe. For example, the neighborhoods of MW and M31 were found in CoDa I to have reionized at roughly the same redshift, a bit earlier than the global average reionization redshift, a result which was also found by \mbox{CoDa I-AMR} as described in \citet[]{codaIamr}. 
\mbox{CoDa I-AMR} used an improved calibration of subgrid star formation efficiency so as to finish global reionization by $z=6.1$, making it easier to relate simulation redshifts directly to observations. In \citet[]{codaIamr}, the reionization redshift field was based on the redshifts when each cell first reached $X_\text{ion}=0.5$ (rather than $X_\text{ion}=0.9$ as used here). As structure formation changes the comoving positions of mass over time, the time when a mass element first reaches a location that is 50\% ionized is defined as its reionization redshift, $z_\textsc{r}$. The average reionization redshift of the mass elements that constitute MW and M31 at $z=0$ were found to be $\langle z_\textsc{r} \rangle=8.2$ for both, while $\langle z_\textsc{r} \rangle_\text{global}=7.8$. The spread of reionization times within each galaxy was as large as $\sim 400$ Myrs \mbox{(max - min)}, with a $2\sigma$ spread of $\sim 100-150$ Myrs.
Our results here suggest that the suppression of the SFRs in the low-mass progenitors of MW and M31 were similar to each other, therefore, but were different from that in regions that reionized much earlier or later.

\section*{Acknowledgements}
This research benefitted from the support of NSF AST-1009799, NASA NNX11AE09G, and SURP/JPL Project Nbr 1492788 and 1515294 to PRS.
The \textsc{ramses-cudaton} simulation, CoDa I, analyzed here was performed on the massively-paralleled Titan supercomputer at Oak Ridge National Laboratory OLCF, under DOE INCITE 2013 award AST031. Post-processing was performed on the Eos, Rhea and Lens clusters. Auxiliary simulations used the PRACE-3IP project (FP7 RI-312763) resource curie-hybrid based in France at Tr\`{e}s Grand Centre de Calcul. The CoDa I-DM2048 \textsc{gadget} simulation was performed at LRZ Munich. The analysis presented here was conducted on the Stampede supercomputer at the Texas Advanced Computing Center of the University of Texas at Austin with NSF XSEDE grant TG-AST090005. PO acknowledges support from the French ANR funded project ORAGE (ANR-14-CE33-0016). NG and DA acknowledge funding from the French ANR for project ANR-12-JS05-0001 (EMMA). ITI was supported by the Science and Technology Facilities Council [grant numbers ST/F002858/1 and ST/I000976/1] and the Southeast Physics Network (SEPNet). SG and YH acknowledge support by DFG grant GO 563/21-1. YH has been partially supported by the Israel Science Foundation (1013/12). GY also acknowledges support from MINECO-FEDER under research grants AYA2012-31101 and AYA2015-63810-P.




\bibliographystyle{mnras}
\bibliography{main} 




\appendix
\section{Identifying the Regions that Imported and Exported Ionizing Radiation During the EOR}
\label{appendix}

It is possible to determine which regions produced more ionizing photons than they consumed and which regions consumed more than they produced, in the process of finishing and maintaining their own local reionization. We do this by counting their relative shares of the total number of ionizing photons produced ($N_{\gamma,i}$, where $N_{\gamma,i} \propto M_{\star,i}$, the total mass of stars formed in region $i$) and recombinations experienced ($N_{\text{rec},i}$) over time to the end of global reionization, and the total number of H atoms they contained ($N_{\textsc{h},i}$), as follows. We shall assume that all of the ionizing photons produced during the EOR were absorbed by the time global reionization ended (or shortly thereafter). In that case, summed over our large simulation volume (or, equivalently, expressed per unit volume, averaged over the universe), we can write
\begin{equation}
    N_{\gamma,\text{tot}} = N_{\text{rec,tot}} + N_{\textsc{h},\text{tot}},
    \label{AllPhotonsAbs}
\end{equation}
where $N_{x,\text{tot}} = \sum_i N_{x,i}$. Any given region $i$ that produced a deficit of ionizing photons, according to the inequality 
\begin{equation}
    N_{\gamma,i} < N_{\text{rec},i} + N_{\textsc{h},i},
    \label{ImportCondition}
\end{equation}
must have been a net `importer' of ionizing photons in order to finish and maintain its own reionization. Conversely, a region that produces a net surplus, according to
\begin{equation}
    N_{\gamma,i} > N_{\text{rec},i} + N_{\textsc{h},i}
    \label{ExportCondition}
\end{equation}
must have been a net `exporter'.

Importers and exporters can be identified simply in terms of $f_{\star,i}$ and $f_{\text{rec},i}$, under reasonable assumptions. If $N_\text{rec,tot} \gg N_{\textsc{h},\text{tot}}$, then regions in which $f_{\star,i}< f_{\text{rec},i}$ were net importers, which we prove as follows. Since $N_{\gamma,i} \propto M_{\star,i}$, $f_{\gamma,i} = f_{\star,i}$. If $f_{\star,i}< f_{\text{rec},i}$, then we can write the following inequalities:
\begin{equation}
    \frac{N_{\gamma,i}}{N_{\gamma,\text{tot}}} < \frac{N_{\text{rec},i}}{N_{\text{rec},\text{tot}}} < \frac{N_{\text{rec},i} + N_{\textsc{h},i}}{N_{\text{rec},\text{tot}}}
    \label{Import1stStep}
\end{equation}
Combining inequality~\ref{Import1stStep} with equation~\ref{AllPhotonsAbs} yields
\begin{equation}
    \frac{N_{\gamma,i}}{N_{\text{rec},\text{tot}} + N_{\textsc{h},\text{tot}}} < \frac{N_{\text{rec},i} + N_{\textsc{h},i}}{N_{\text{rec},\text{tot}}}
\end{equation}
If $N_\text{rec,tot} \gg N_{\textsc{h},\text{tot}}$, this becomes
\begin{equation}
    N_{\gamma,i} < N_{\text{rec},i} + N_{\textsc{h},i} + \mathcal{O}\Big(\frac{N_{\textsc{h},\text{tot}}}{N_{\text{rec},\text{tot}}}\Big),
\end{equation}
so region $i$ was a net importer according to inequality~\ref{ImportCondition}.
Conversely, if $N_{\textsc{h},i} \not\gg N_{\text{rec},i}$, then region $i$ was a net exporter if $f_{\star,i}\gg f_{\text{rec},i}$, which we prove as follows. Since
\begin{equation}
    \frac{N_{\gamma,i}}{N_{\gamma,\text{tot}}} \gg \frac{N_{\text{rec},i}}{N_{\text{rec},\text{tot}}} > \frac{N_{\text{rec},i}}{N_{\text{rec},\text{tot}} + N_{\textsc{h},\text{tot}}},
    \label{Export1stStep}
\end{equation}
then using equation~\ref{AllPhotonsAbs} to replace $N_{\gamma,\text{tot}}$ on the left-hand-side of inequality~\ref{Export1stStep} yields 
\begin{equation}
    N_{\gamma,i} \gg N_{\text{rec},i}
\end{equation}
Inequality~\ref{ExportCondition} follows if we assume that $N_{\textsc{h},i} \not\gg N_{\text{rec},i}$.

Our assumption that $N_\text{rec,tot} \gg N_{\textsc{h},\text{tot}}$ is easily justified as follows. Per unit volume, the integrated number of recombinations through the end of global reionization at $z_\text{end}$ is given by
\begin{equation}
    \langle n_\text{rec} \rangle \equiv \frac{N_\text{rec,tot}}{V} = \int_{z_\text{begin}}^{z_\text{end}}\mathcal{R}\frac{dt}{dz}dz,
    \label{n_rec,tot}
\end{equation}
where $\mathcal{R}$ is given by equation~\ref{recombrate} and $z_\text{begin}$ is some redshift before reionization begins, post-recombination. The cell-wise results of CoDa I can be used to evaluate equation~\ref{n_rec,tot}. For $z_\text{end}=4.6$, we find that $N_\text{rec,tot}/N_{\textsc{h},\text{tot}} = \langle n_\text{rec}\rangle/\langle n_{\textsc{h}}\rangle \simeq 5.5\times10^{-4} \text{ cm}^{-3}/3.3\times10^{-5}\text{ cm}^{-3} \approx 20$, so $N_\text{rec,tot} \gg N_{\textsc{h},\text{tot}}$, as required\footnote{
This is based upon smoothing the original CoDa I grid of $4096^3$ cells onto $2048^3$ cells. The unsmoothed data would have yielded a higher number of recombinations, making the inequality even stronger.}. 

This is consistent with an estimate of the ratio of the mean recombination time per atom to the Hubble time during the EOR given by
\begin{equation}
    \frac{t_\text{rec}}{t_\textsc{h}} = \frac{H(a)}{\langle \alpha(T) n_e(a) c_l(a) \rangle},
\end{equation}
where $c_l$ is the ionized gas density clumping factor $\langle n_e^2 \rangle/\langle n_e \rangle^2$. For the matter-dominated flat universe at early times, when $\Lambda$ is unimportant, this is given by
\begin{equation}
    \frac{t_\text{rec}}{t_\textsc{h}} \simeq \frac{H_0 \Omega_m^{1/2} a^{3/2}}{\langle \alpha(T) n_{\textsc{h},0} X_\text{ion} c_l(a) \rangle} \simeq c_l^{-1}X_\text{ion}^{-1}\Big(\frac{1+z}{6}\Big)^{-3/2},
\end{equation}
assuming photoionized gas has $T\sim10^4$ K. Since $c_l \gg 1$ at late times during the EOR, when averaging over the entire volume, including that inside halos, $t_\text{rec}/t_\textsc{h} \ll 1$ for $X_\text{ion}\sim1$.
[Note: The fact that $N_\text{rec,tot}/N_{\textsc{h},\text{tot}} \gg 1$ is not in conflict with estimates of the average number of ionizing photons per atom required to finish global reionization of the IGM, which is typically a much smaller number, of order a few or less. This is because the latter, smaller number is only counting the number of ionizing photons that {\it escape} from galaxies into the IGM, while $N_\text{rec,tot}$ accounts for all those additional ionizing photons {\it produced} inside galaxies which do {\it not} escape, as well.]

We also need to justify the second condition, that $N_{\textsc{h},i} \not\gg N_{\text{rec},i}$ for the case of exporter regions. Our CoDa I results show that these are the early-reionizing regions, which are also the most overdense. As such, if $N_\text{rec,tot} \gg N_{\textsc{h},\text{tot}}$ over all, then so must $N_{\text{rec},i} \gg N_{\textsc{h},i}$ locally for these regions.


\bsp	
\label{lastpage}
\end{document}